\documentclass[submit]{epsv8}
\usepackage{lineno}
\usepackage{amssymb}
\usepackage{amsmath}
\usepackage{color}
\usepackage{graphicx}

\newcommand {\reviser}[1]{\textcolor{black}{#1}}
\newcommand {\reviseb}[1]{\textcolor{black}{#1}}
\newcommand {\revrr}[1]{\textcolor{black}{#1}}
\newcommand {\vect}[1]{\mbox{\boldmath $#1$}}

\newcommand {\inty}[2]{\int_{#1}^{#2}}

\newcommand {\pdif}[3][]{\frac{\partial^{#1}#2}{\partial#3^{#1}}}

\newcommand {\lsim}{\hspace{0.3em}\raisebox{0.4ex}{$<$}\hspace{-0.75em}\raisebox{-.7ex}{$\sim$}\hspace{0.3em}}

\makeatletter
\def\mart{\@ifnextchar[{\mart@@}{\mart@}}
\def\mart@@[#1]#2{\sqrt[#1]{\mathstrut{#2}}}
\def\mart@#1{\sqrt{\mathstrut{#1}}}
\makeatother
\newcommand {\Alfven}{Alfv\'{e}n}

\title{A study of solar energetic particle transport on 30 March 2022 using multi-spacecraft data assimilation}
\author{
 Takashi Minoshima (corresponding author), Center for Mathematical Science and Advanced Technology, Japan Agency for Marine-Earth Science and Technology, 3173-25 Syowa-machi, Kanazawaku, Yokohama 236-0001, Japan, minoshim@jamstec.go.jp\\
 Yoshizumi Miyoshi, Institute for Space-Earth Environmental Research, Nagoya University, Furo-cho, Chikusa-ku, Nagoya 464-8601, Japan; Kyung Hee University, Swon, Korea, miyoshi@isee.nagoya-u.ac.jp\\
 Go Murakami, Institute of Space and Astronautical Science, Japan Aerospace Exploration Agency, 3-1-1 Yoshinodai, Chuo-ku, Sagamihara, Kanagawa 252-5210, Japan, go@stp.isas.jaxa.jp\\
 Marco Pinto, European Space Research and Technology Centre, European Space Agency, Keplerlaan 1, Postbus 299, 2200 AG Noordwijk, Netherlands; Laboratory of Instrumentation and Experimental Particle Physics, Av. Prof. Gama Pinto, n.2, 649-003 Lisboa, Portugal, mgpinto11@gmail.com\\
 Daniel Schmid, Space Research Institute, Austrian Academy of Sciences, Schmiedlstra\ss e 6 A-8042 Graz, Austria, Daniel.Schmid@oeaw.ac.at\\
 Ayako Matsuoka, Data Analysis Center for Geomagnetism and Space Magnetism, Graduate School of Science, Kyoto University, Kitashirakawa-Oiwake Cho, Sakyo-ku, Kyoto 606-8502, Japan, matsuoka@kugi.kyoto-u.ac.jp\\
 Wolfgang Baumjohann, Space Research Institute, Austrian Academy of Sciences, Schmiedlstra\ss e 6 A-8042 Graz, Austria, Wolfgang.Baumjohann@oeaw.ac.at\\
 David Fischer, Space Research Institute, Austrian Academy of Sciences, Schmiedlstra\ss e 6 A-8042 Graz, Austria, david.fischer@oeaw.ac.at\\
 Kazumasa Iwai, Institute for Space-Earth Environmental Research, Nagoya University, Furo-cho, Chikusa-ku, Nagoya 464-8601, Japan, k.iwai@isee.nagoya-u.ac.jp\\
 Shinsuke Imada, Department of Earth and Planetary Science, Graduate School of Science, The University of Tokyo, 7-3-1 Hongo, Bunkyo-ku, Tokyo 113-0033, Japan, imada@eps.s.u-tokyo.ac.jp\\
}

\abstract{
We analyze a \reviseb{unique} solar energetic particle event observed \reviseb{simultaneously} by the BepiColombo and STEREO-A spacecraft on March 30, 2022.
The two spacecraft at heliocentric distances of 0.6 and 1.0 AU are expected to be aligned approximately along the same magnetic field line, providing a valuable opportunity to investigate particle transport processes \reviseb{ in the inner heliosphere.}
Protons with energies above 1.0 MeV exhibit velocity dispersion during the rise phase, suggesting that the energetic particles are produced close to the Sun, possibly associated with a coronal mass ejection.
In contrast, protons during the decay phase are characterized by long-lasting time profiles with longer time scales at 1.0 AU than at 0.6 AU, suggesting that the particles deviate from ballistic propagation.
\reviseb{By assimilating these multi-spacecraft observation data into numerical simulations of the focused transport equation, for the first time, we estimate the mean free path parallel to the magnetic field as a time series.}
The inferred mean free path decreases over time and approaches around 0.5-1.0 AU at the STEREO-A location during the decay phase, suggesting an increasing influence of scattering on particle transport.
This interpretation is qualitatively supported by independent STEREO-A observations that showed increasing magnetic field fluctuations, \reviseb{suggesting the connection between the particle transport and the local field fluctuations.}
However, only a fraction of these fluctuations is expected to contribute to particle scattering, \reviseb{which may be due to the multidimensional nature of magnetic field fluctuations.}
}


\begin{document}

\maketitle

\section{Introduction}\label{sec:introduction}
Solar Energetic Particles (SEPs) are high-energy charged particles ranging from a few keV to several GeV, produced by energetic phenomena on the Sun and subsequently ejected into interplanetary space \citep{1999SSRv...90..413R,2016LRSP...13....3D}. 
\reviseb{SEP events offer a unique natural laboratory for investigating particle acceleration and wave-particle interaction processes in dilute plasmas through in-situ observations. 
A deeper understanding of the origin and dynamics of SEPs is therefore fundamental not only to space plasma physics, but also to astrophysics and related research fields.}
Furthermore, accurate prediction of the SEP profile is crucial for space weather operations, as SEPs above 10 MeV pose primary threats to modern society, causing radio communication failures, malfunction and degradation of equipment onboard aircraft and satellites, and radiation exposure of astronauts \citep[e.g.,][]{2021GMS...262...63T,2021GMS...262...79M}. 
This is expected to grow in importance as human activities expand beyond the Earth's magnetosphere.

The motion of energetic particles in space plasmas can be characterized by the mean free path, which is governed primarily by wave-particle interactions rather than binary collisions.
A higher wave intensity leads to a shorter mean free path due to more frequent interactions, enhancing the particle confinement.
A notable application is diffusive shock acceleration of particles, where the particle flux increases exponentially from the upstream to the downstream side of the shock \citep{1978MNRAS.182..147B,1978ApJ...221L..29B}. 
The mean free path (or equivalently, the diffusion coefficient) not only controls the spatial profile of particles but also the efficiency of particle acceleration and the maximum attainable energy.
Accurately determining the mean free path is therefore crucial for understanding the acceleration and transport of energetic particles, including SEPs \citep[e.g.,][]{2000JGR...10525079Z,2012AdSpR..49.1067L,2017JGRA..12210938H}.

Once SEPs are released from their acceleration sites of solar flares and coronal mass ejections (CMEs), they propagate primarily along the nominal Parker spiral magnetic field. 
Depending on solar wind conditions, SEPs travel either ballistically or diffusively, both along and across the magnetic field lines.
These processes are characterized by the parallel and perpendicular mean free paths, $\lambda_{\parallel}$ and $\lambda_{\perp}$.
Accurate knowledge of the mean free path is essential for reliable SEP predictions.
A comparative study on estimating the mean free path from SEP observations was summarized by \cite{1982RvGSP..20..335P}, who reported that the parallel mean free path measured at 1.0 AU ranges from 0.08 to 0.3 AU across a broad range of particle rigidity ($5\times 10^{-4}$ to 5 GV).
However, this consensus conflicts with predictions by standard quasi-linear theory \citep{1966ApJ...146..480J}, both in magnitude and rigidity dependence.
Subsequent researches have focused on reconciling this discrepancy between theory and observation \citep[e.g.,][]{1993JGR....98.3513W,1993AdSpR..13i.359W}.
In particular, \cite{1994ApJ...420..294B} proposed incorporating the dissipation range and the dynamic nature of magnetic field fluctuations to modify the rigidity dependence derived from the standard theory, thereby achieving better agreement with observations.
Additionally, \cite{1996JGR...101.2511B} considered the multidimensional nature of magnetic field fluctuations to address the issue that the mean free path derived from the standard theory is too short compared to that from SEP observations.
These theoretical advancements have been verified through comparisons between observation and numerical modeling \citep{2003ApJ...589.1027D,2009ApJ...693...69D,2013JGRA..118..642T,2024ApJ...971..105L}.

To estimate the mean free path from SEP observations, time profiles of particle flux and first-order anisotropy are modeled by numerical solutions of the focused transport equation, which treats the parallel mean free path as a fitting parameter \cite[e.g.,][]{2000ApJ...537.1073D,2005ApJ...627..562Q,2006JGRA..111.8101Q}.
When observational data are available from only a single location, it is necessary to model the injection time profile at a reference location to solve the equation.
Delta function or a phenomenological profile is frequently employed, and they are also determined by fitting.
However, because the mean free path is closely tied to the time profile, uncertainties in the injection profile significantly affect the mean free path estimation.
This limitation can be mitigated by using simultaneous observations at multiple magnetically well-connected locations.
At present, many in-situ spacecraft including Parker Solar Probe, Solar Orbiter, BepiColombo, ACE, STEREO, and JUICE are capable of measuring SEPs at various heliocentric distances and longitudes.
By integrating simultaneous observations from various locations with numerical models, one expects to get a better understanding of acceleration and transport processes of SEPs.
For example, \cite{2014JGRA..119.6074D,2016ApJ...826..134D} analyzed data taken by STEREO and near-Earth spacecraft at different longitudes in conjunction with a three-dimensional numerical model to investigate the parallel and perpendicular mean free paths of energetic electrons.
\cite{2024JSWSC..14....3P} conducted an integrated study using multi-spacecraft observations and scatter-free focused transport simulations for energetic protons, discussing the necessity of particle acceleration within the solar corona.

In this paper, we report on a unique SEP event that occurred on March 30, 2022, observed simultaneously by the BepiColombo and STEREO-A spacecraft at heliocentric distances of 0.6 and 1.0 AU, respectively.
The two spacecraft were approximately aligned along the same magnetic field line, providing a valuable opportunity to investigate SEP transport between magnetically-connected observation points.
Using this configuration, we apply a data assimilation technique to numerical simulations of the focused transport equation for the first time in multi-spacecraft SEP observations.
By assimilating the observed SEP flux at 1.0 AU into a model driven by the observation at 0.6 AU, we aim to infer the parallel mean free path as a time series.
In \S~\ref{sec:observation}, we report the SEP observation along with the corresponding solar wind condition.
In \S~\ref{sec:numerical-model}, we perform the data assimilation into the focused transport simulations to interpret the observation.
Finally in \S \ref{sec:dis}, we discuss a possible scenario for this SEP event based on the observation and simulation results.

\section{Observation}\label{sec:observation}
\begin{figure}[htbp]
\centering
\includegraphics[clip,angle=0,scale=0.3]{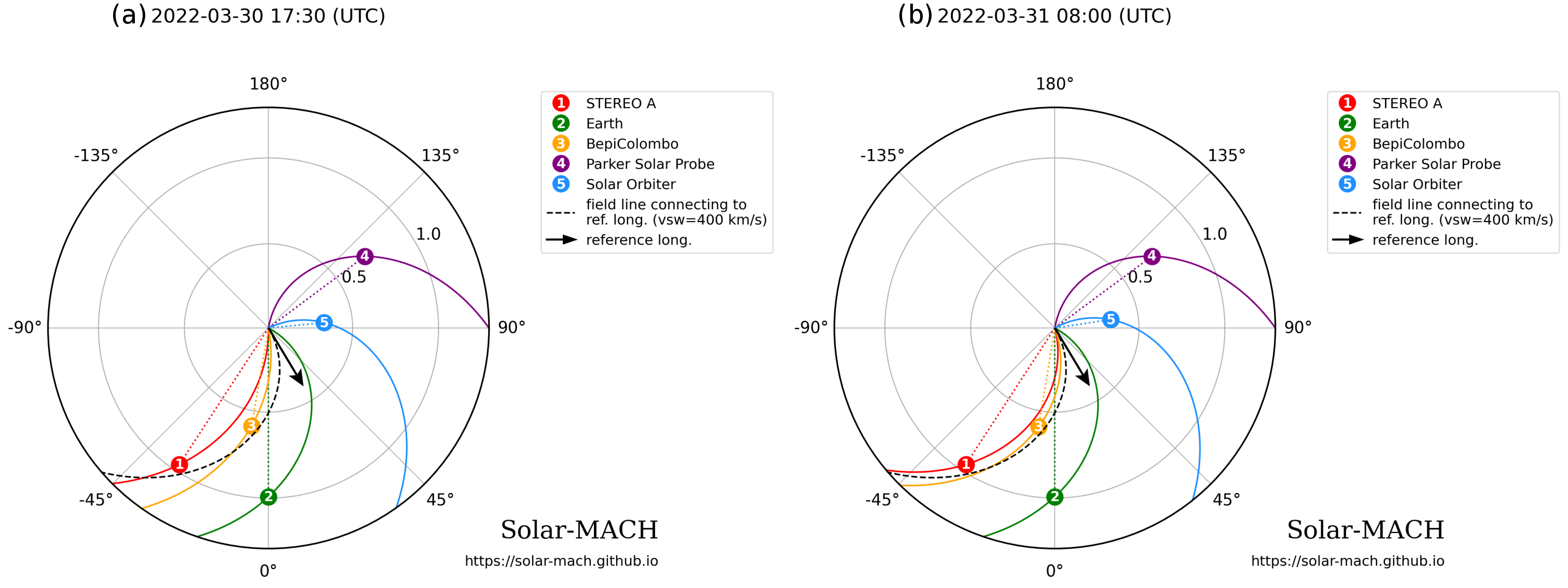}
\caption{Spacecraft location and magnetic field connection at \reviser{(a) 17:30 UT on March 30, 2022, and (b) 08:00 UT on March 31, 2022}, created by the Solar-Mach Python tool \citep{2023FrASS...958810G}. The flare and its magnetic connection are denoted as the black arrow and the dashed line, respectively. Solar wind speeds of 700 ${\rm km \; s^{-1}}$ \reviser{in (a) and 500 ${\rm km \; s^{-1}}$ in (b)} are used at the BepiColombo and STEREO-A locations, and 400 ${\rm km \; s^{-1}}$ elsewhere.}
\label{fig:solarmach} 
\end{figure}

An intense X1.3 class flare occurred at N13W31 (AR12975) at 17:21 UT on March 30, 2022.
According to the SOHO/LASCO CME Catalog \citep{2009EM&P..104..295G}, a westward-directed CME was subsequently observed at 18:00 UT with speed of $641 \; {\rm km \; s^{-1}}$.
At that time, the BepiColombo and STEREO-A spacecraft were located at heliocentric distances of 0.6AU and 1.0AU, respectively.
Figure \ref{fig:solarmach} shows the inferred spacecraft locations created by the Solar-Mach Python tool \citep{2023FrASS...958810G}.
\reviser{Since the solar wind speed at the STEREO-A location decreases over time (as shown in \S~\ref{sec:solar-wind}), we present two images corresponding to different speeds at STEREO-A: (a) $700 \; {\rm km \; s^{-1}}$ at 17:30 UT on March 30,  and (b) $500 \; {\rm km \; s^{-1}}$ at 08:00 UT on March 31. 
We assume that the solar wind speed at the BepiColombo location is the same as that at STEREO-A, because plasma data is not available for this event.}
The Stonyhurst heliographic coordinates were $(-1.6^{\circ}, -9.8^{\circ}\sim-9.3^{\circ})$ for BepiColombo and $(-7.3^{\circ},-33.1^{\circ})$ for STEREO-A. 
\reviseb{Solar-Mach estimates their magnetic footpoint longitudes on the solar surface to be $11.1^{\circ}-20.2^{\circ}$ for BepiColombo and $1.0^{\circ}-14.6^{\circ}$ for STEREO-A. 
\cite{2024Sci...385..962R} reported simultaneous observations of the plasma stream in February 2022 by Parker Solar Probe at 0.06 AU and Solar Orbiter at 0.6 AU, with a corresponding longitudinal spread on the source surface being $5^{\circ}$. 
Although such conditions can vary from event to event, this result may support our assumption that BepiColombo and STEREO-A were aligned approximately along the same magnetic field line and connected to a similar source region within a longitudinal spread of about $5^{\circ} - 10^{\circ}$. }

\subsection{Energetic particles}\label{sec:energetic-particles}
During this event, energetic particles were detected by the BepiColombo Environment Radiation Monitor \citep[BERM;][]{2022SSRv..218...54P} onboard the Mercury Planetary Orbiter \citep[MPO;][]{2021SSRv..217...90B}, Solar Particle Monitor (SPM) onboard the Mercury Magnetospheric Orbiter \citep[Mio;][]{2020SSRv..216..113M,2025JGRA..13033147K}, and the Low Energy Telescope \cite[LET;][]{2008SSRv..136..285M} onboard STEREO-A.
Owing to their locations relative to the Parker spiral field, both spacecraft are expected to have observed the same particle population at different heliocentric distances. 

\begin{figure}[htbp]
\centering
\includegraphics[clip,angle=0,scale=0.7]{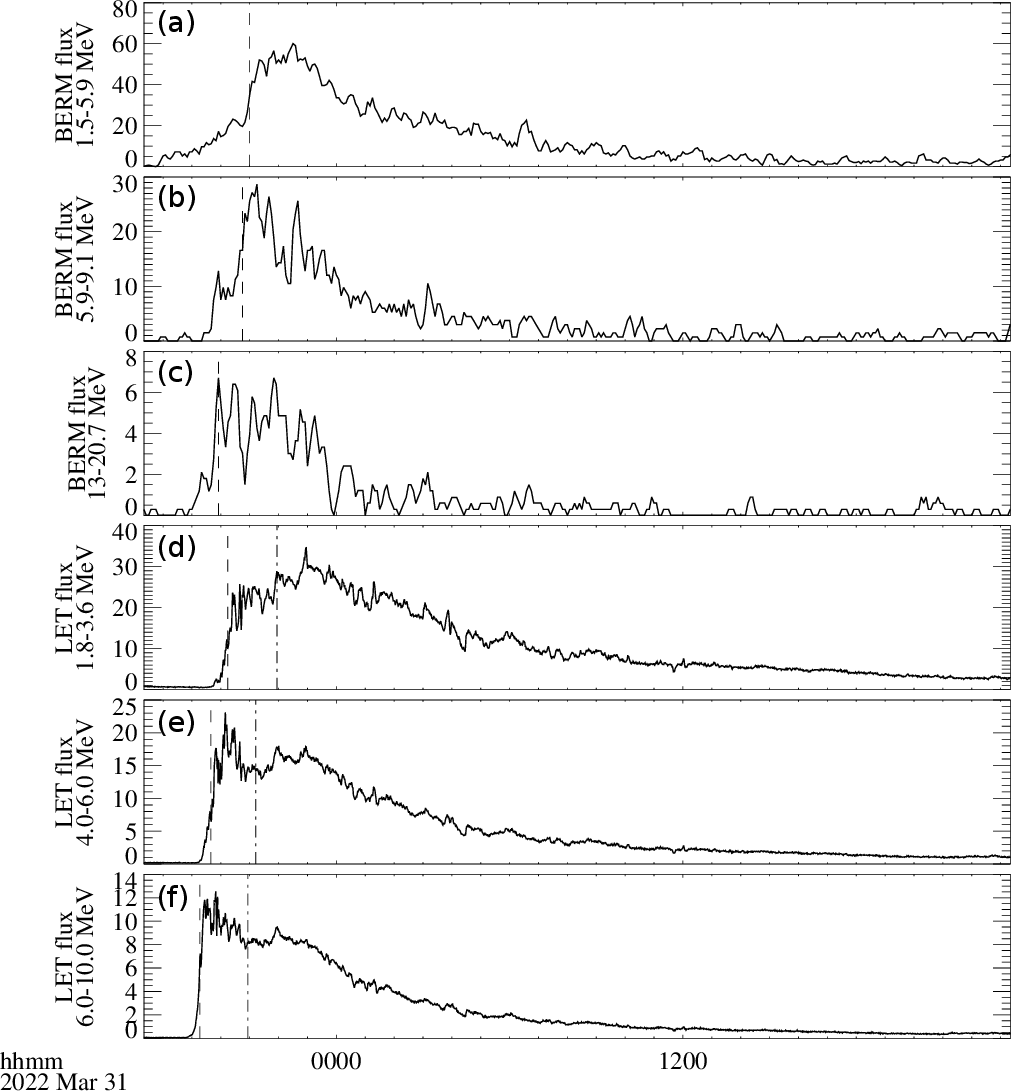}
 \caption{Time profiles of the SEP differential flux (particles cm$^{-2}$ s$^{-1}$ sr$^{-1}$ MeV$^{-1}$) during 2022/03/30 17:20 UT - 2022/03/31 23:20 UT. (a-c) 600 seconds cadence data taken by MPO/BERM at 0.6 AU in 1.5-5.9 MeV, 5.9-9.1 MeV, and 13.0-20.7 MeV. (d-f) 60 seconds cadence data taken by STEREO-A/LET at 1.0 AU in 1.8-3.6 MeV, 4.0-6.0 MeV, and 6.0-10.0 MeV. Vertical dashed lines indicate the inferred onset time at each channel. \reviser{Vertical dashed-dotted lines in the LET profile indicate the onset time of the long-lasting component estimated from the velocity dispersion analysis of the BERM time profile.}}
\label{fig:berm_let_prof}
\end{figure}

Figure \ref{fig:berm_let_prof} shows the proton fluxes observed by BERM at 0.6 AU and LET at 1.0 AU in three energy channels over 30 hours from 17:20 UT on March 30 to 23:20 UT on March 31.
At 0.6 AU, the profiles exhibit rapid rise until 23:00 UT, followed by gradual decay.
The BERM fluxes peak at 22:30 UT for 1.5-5.9 MeV, 21:15 UT for 5.9-9.1 MeV, and 19:55 UT for 13.0-20.7 MeV, exhibiting clear velocity dispersion.
In contrast, two distinct components are identified in the LET profiles at 1.0 AU: a preceding component characterized by a \reviser{sharp} rise with high fluctuation until 21:00 UT, and a subsequent long-lasting component.
The preceding components peak at 20:40 UT for 1.8-3.6 MeV, 20:10 UT for 4.0-6.0 MeV, and 19:50 UT for 6.0-10.0 MeV, exhibiting velocity dispersion.
The long-lasting components peak at 22:57 UT for 1.8-3.6 MeV, 22:00 UT for 4.0-6.0, and 21:57 UT for 6.0-10.0 MeV.
\revrr{It is expected that the 6.0 MeV preceding component would peak around 19:20 UT at 0.6 AU. 
However, the 5.9-9.1 MeV BERM flux shown in panel (b) does not exhibit a clear peak at this time.
This suggests that the preceding component detected by LET was not observed by BERM.}

\begin{figure}
\centering
\includegraphics[clip,angle=0,scale=0.7]{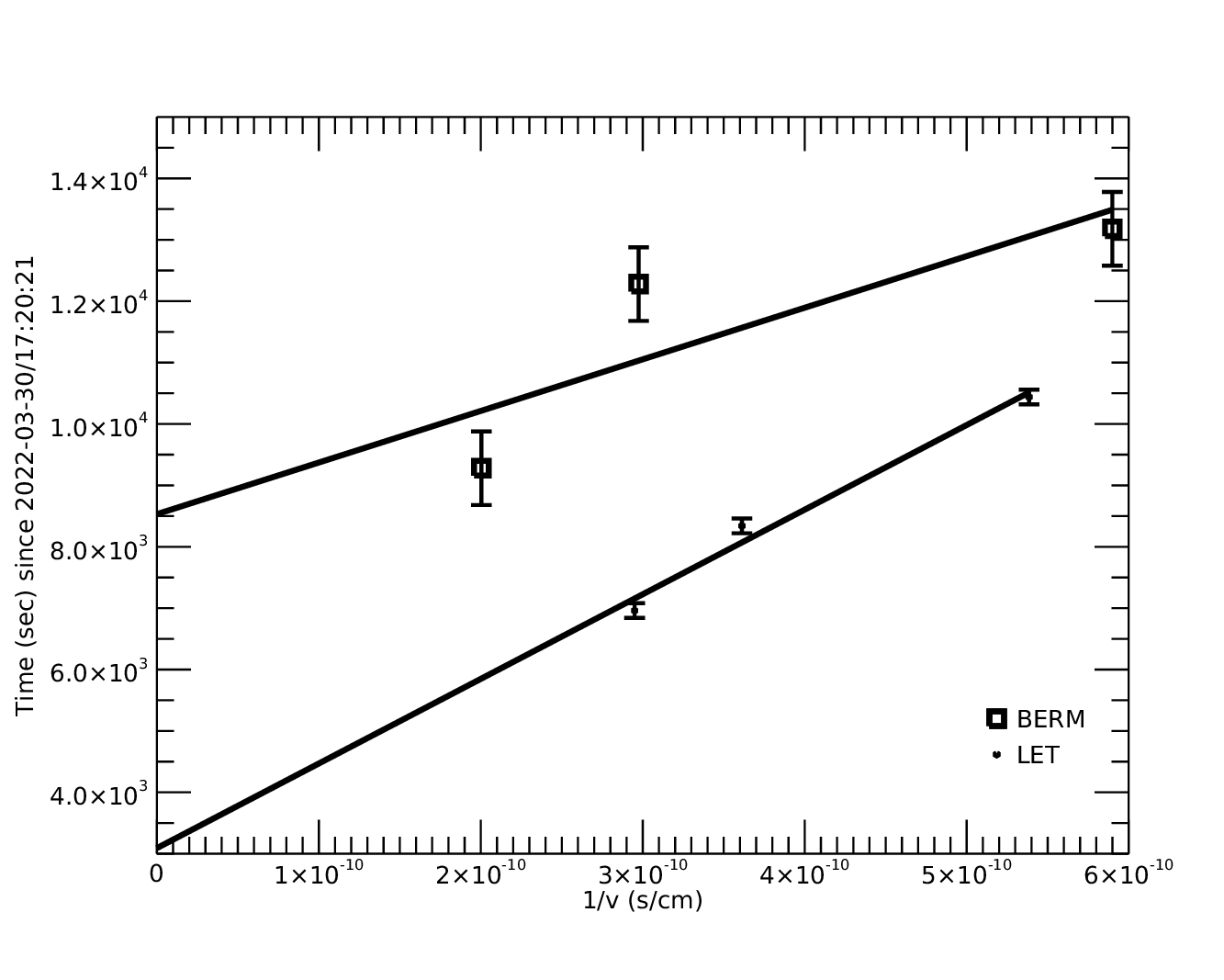}
\caption{Velocity dispersion analysis of the MPO/BERM (plus) and STEREO-A/LET (asterisk) data. \reviseb{The error bars indicate the temporal resolution of the instruments.} Solid lines represent the regression line for each instrument.}
\label{fig:tof_summary}
\end{figure}

The velocity-dependent peak times in BERM and LET suggest that the particles followed ballistic propagation at least during the rise phase.
To estimate the energy-independent release time $t_0$ and the path length $L$ from the release site, we conduct a velocity dispersion analysis \cite[e.g.,][]{2004A&A...420..343L}.
Using the time profiles of the BERM 1.5-5.9 MeV and LET 6.0-10.0 MeV energy channels, we first define the onset time as the moment when the flux exceeds half the maximum value.
We then calculate the time lags between adjacent energy channels using the cross correlation of the time profiles in the rise phase, yielding the onset time for each energy channel, $t_{\text on}(E)$, as indicated by vertical dashed lines in Figure \ref{fig:berm_let_prof}.
The release time and the path length are inferred by fitting a regression line to data, $t_{\text on}(E) = t_0 + L/v(E)$, where $v(E)$ is the particle speed.
The result is shown in Figure \ref{fig:tof_summary}.
The inferred path lengths are $0.56\pm0.07$ AU from the BepiColombo location and $0.92\pm0.02$ AU from the STEREO-A location (the error is evaluated from the temporal resolution of each instrument).
\revrr{Using a solar wind velocity of $700 \; {\rm km \; s^{-1}}$, the pass lengths from the Sun to the two instruments are 0.608 AU and 1.05 AU, respectively. The heliocentric distance of the particle release site is thus inferred to be 0.05-0.13 AU.}
However, the inferred release time from BepiColombo $\sim$ 19:42 UT is much later than that from STEREO-A $\sim$ 18:12 UT, indicating possible differences in particle population observed at the two spacecraft during the rise phase, \reviser{which is discussed in \S~\ref{sec:dis}.}

\reviser{In Figure \ref{fig:tof_summary}, the velocity dispersion analysis of the LET time profile is applied to the preceding component.
The onset time of the long-lasting component in the LET time profile is inferred from the travel time over 1.0 AU, assuming an injection at 19:42 UT as estimated from the velocity dispersion analysis of the BERM time profile.
The expected onset times are 21:57 UT, 21:13 UT, and 20:56 UT for 1.8 MeV, 4.0 MeV, and 6.0 MeV, respectively, as indicated by vertical dash-dotted lines in Figure \ref{fig:berm_let_prof}.
The good agreement between these expected onsets and the LET time profile suggests that the long-lasting component observed by the both instruments originates from the same population.}

In contrast to the rapid rise, both BERM and LET observed gradually decaying profiles.
We fit these profiles using a commonly adopted phenomenological function \citep[e.g.,][]{2000ApJ...537.1073D},
\begin{eqnarray}
f_{\rm fit}(t)=\frac{C}{t}\exp\left(-\frac{\tau_r}{t}-\frac{t}{\tau_d}\right), \label{eq:12}
\end{eqnarray}
where $\tau_r$ and $\tau_d$ denote the rise and decay time scales, respectively.
The fitted decay time scales increase with distance: at the BepiColombo location, they are 7.2 hours for 1.5-5.9 MeV and 3.5 hours for 5.9-9.1 MeV;  at the STEREO-A location, they elongate to 23 hours for 1.8-3.6 MeV, 19 hours for 4.0-6.0 MeV, and 13 hours for 6.0-10.0 MeV.
Assuming that both spacecraft observed the same population \reviser{for the long-lasting component}, these results suggest that the particle transport during the decay phase deviates from ballistic propagation.

\begin{figure}
\centering
\includegraphics[clip,angle=0,scale=0.7]{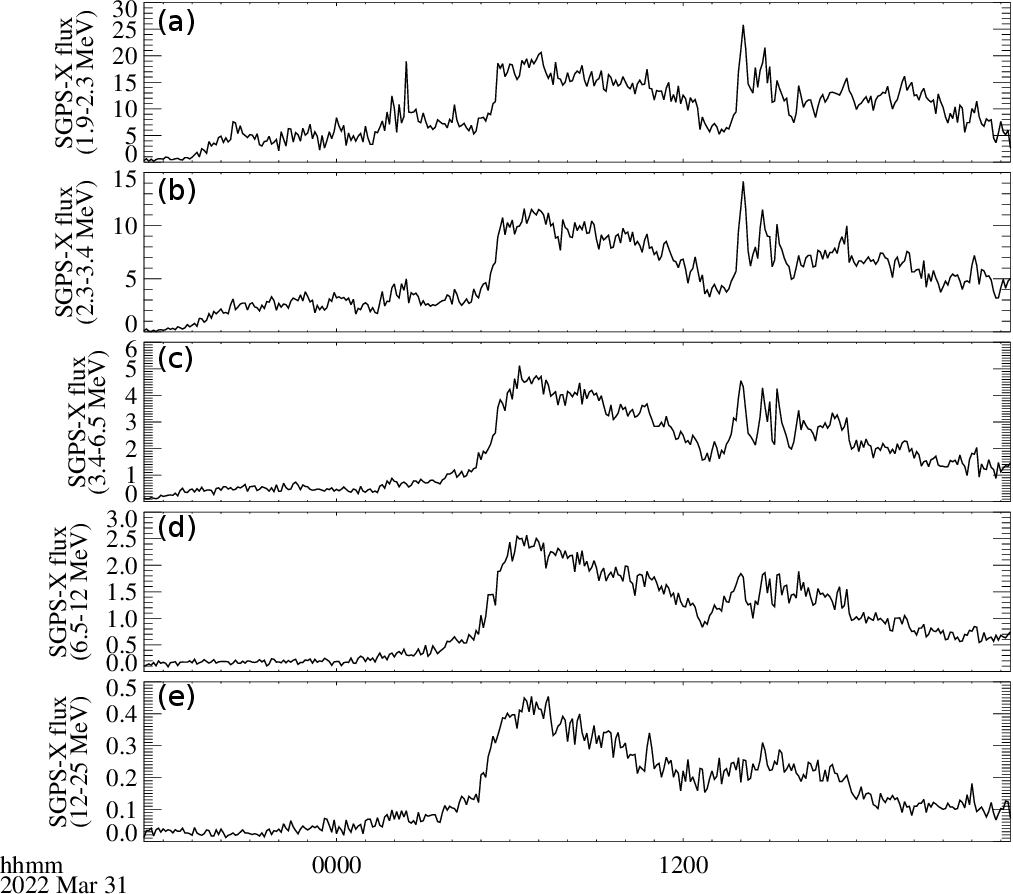}
\caption{Time profiles of the SEP differential fluxes (particles cm$^{-2}$ s$^{-1}$ sr$^{-1}$ MeV$^{-1}$) taken by the GOES-16 SGPS-X during 2022/03/30 17:20 UT - 2022/03/31 23:20 UT. (a-e) 300 seconds cadence data in 1.9-2.3 MeV, 2.3-3.4 MeV, 3.4-6.5 MeV, 6.5-12 MeV, and 12-25 MeV.} 
\label{fig:goes_profile} 
\end{figure}

Figure \ref{fig:goes_profile} shows the 1.9-25 MeV proton fluxes at the Earth's orbit, observed by the Solar and Galactic Proton Sensor \citep[SGPS;][]{2021SpWea..1902750K} onboard GOES during the same period.
At Earth, the fluxes began to increase around 05:00 UT on March 31 and exhibited two distinct peaks at approximately 07:00 UT and 14:00 UT.
This enhancement originated from the arrival of the CME associated with a solar flare \revrr{that} occurred on March 28, and is outside the scope of this study.
Importantly, the fluxes with energies above 3.4 MeV did not show significant enhancement during the period corresponding to the main phase of the SEP profile at STEREO-A.
A similar temporal behavior was seen in the 1.9-4.8 MeV proton flux taken by ACE/EPAM, and the 4.3-7.8 MeV flux by SOHO/COSTEP (although some data gaps are present).

\subsection{Solar wind}\label{sec:solar-wind}
\begin{figure}
\centering
 \includegraphics[clip,angle=0,scale=0.7]{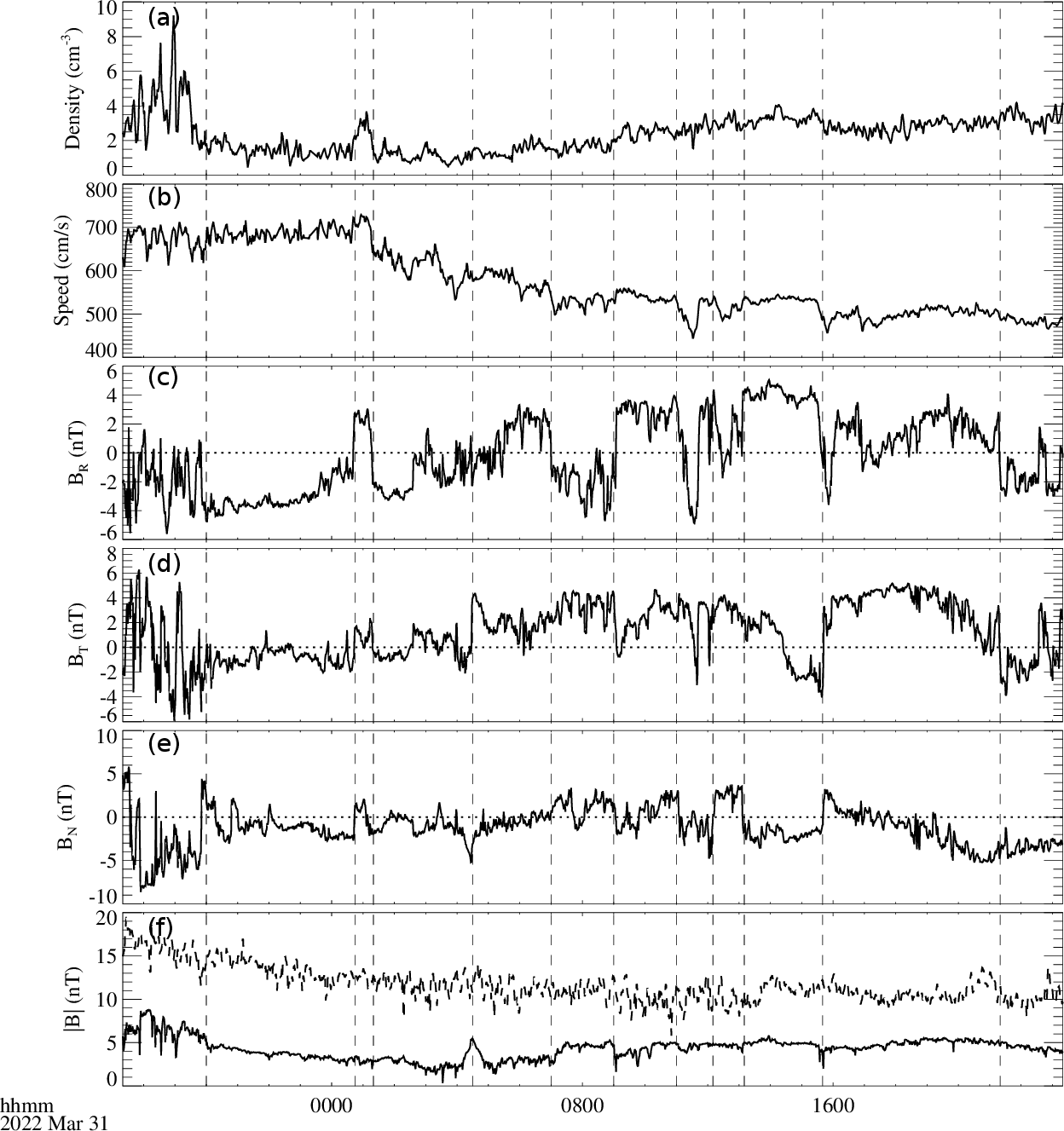}
\caption{\reviser{Time profiles of 60 seconds cadence (a-b) solar wind density and velocity taken by STEREO-A/PLASTIC at 1.0 AU, (c-e) magnetic field vector $(B_R, B_T, B_N)$ taken by STEREO-A/IMPACT, and (f) total magnetic field strength taken by IMPACT (solid line) and Mio/MGF at 0.6 AU (dashed line). Time windows for the Fourier analysis are indicated by vertical dashed lines.}}
\label{fig:sta_pla_mag}
\end{figure}

Figure \ref{fig:sta_pla_mag} shows the time profiles of the plasma and magnetic field data during this event taken by the PLASTIC and IMPACT instruments onboard STEREO-A \reviser{\citep{2008SSRv..136..437G,2008SSRv..136..117L}}, which monitored the solar wind condition at 1.0 AU.
\reviser{In panel (f), the total magnetic field strength is compared between the BepiColombo and STEREO-A locations, with the former taken by the fluxgate magnetometer MGF onboard Mio \citep{2020SSRv..216..125B}.}
Shortly after the event onset at 17:20 UT on March 30, \reviser{STEREO-A} \reviser{was located in the fast solar wind, whose speed is $600-700 \; {\rm km \; s^{-1}}$.}
Both density and the magnetic field exhibited high fluctuations.
These fluctuations ceased by 20:00 UT (the leftmost dashed line).
Subsequently, the solar wind conditions remained relatively stable, with a density of $1.5 \; {\rm cm^{-3}}$ and a speed of $685 \; {\rm km \; s^{-1}}$.
The magnetic field components during this period were $(B_R,B_T,B_N) = (-3.1,-0.8,-1.1)$, until the spacecraft encountered a discontinuity at 00:40 UT on March 31.
This relatively stable period coincides with the SEP rise phase.

The solar wind condition changed significantly after 00:40 UT on March 31.
The density increased to $3.0 \; {\rm cm^{-3}}$ and the velocity decreased to $500 \; {\rm km \; s^{-1}}$ by 23:20 UT.
During this period, the spacecraft encountered multiple magnetic field discontinuities, and the mean field direction varied frequently.
Remarkable changes occurred between 02:30 UT to 11:00 UT.
From 02:30 UT to 07:00 UT, the $B_R$ component changed from negative to positive values, while the $B_T$ and $B_N$ components remained mostly positive and negative, respectively.
The mean field vector during this period was $(B_R,B_T,B_N) = (0.2,1.2,-1.0)$. 
After 07:00 UT, the $B_R$ component reversed back to negative $\simeq -2.0$, and then again to positive $\simeq 2.7$ at 09:00 UT.
During this period, the $B_T$ component remained positive $\simeq 2.8$, while the $B_N$ component frequently changed the direction.
Moreover, the amplitude of magnetic field fluctuations during this period was much higher than in the earlier period.
These large magnetic field variations indicate a considerable deviation from the nominal Parker spiral field, which can strongly influence the particle transport during the decay phase.

\begin{figure}
\centering
 \includegraphics[clip,angle=0,scale=0.7]{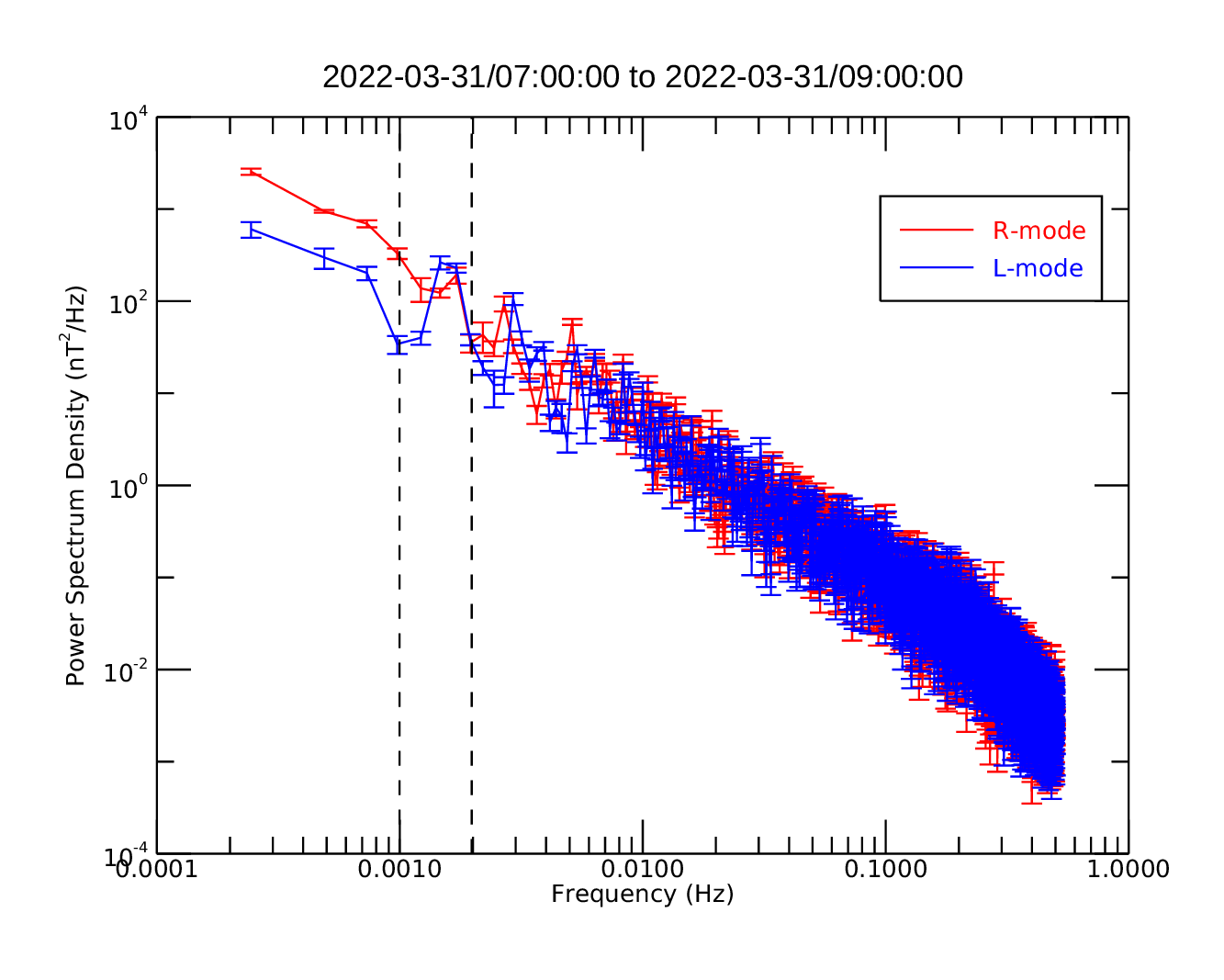}
\caption{Power spectrum density of the magnetic field fluctuation \reviser{taken by STEREO-A} during 2022/03/31 07:00 UT - 2022/03/31 09:00 UT. The red and blue lines correspond to the R- and L-mode components, respectively. \reviseb{The error bars indicate the 95\% confidence interval estimated from 50 samples during this period.} The vertical dashed lines represent the resonance frequency of 1.5 MeV and 5.9 MeV protons (Equation (\ref{eq:7})).}
\label{fig:bpower_plot} 
\end{figure}

\begin{figure}
\centering
 \includegraphics[clip,angle=0,scale=0.7]{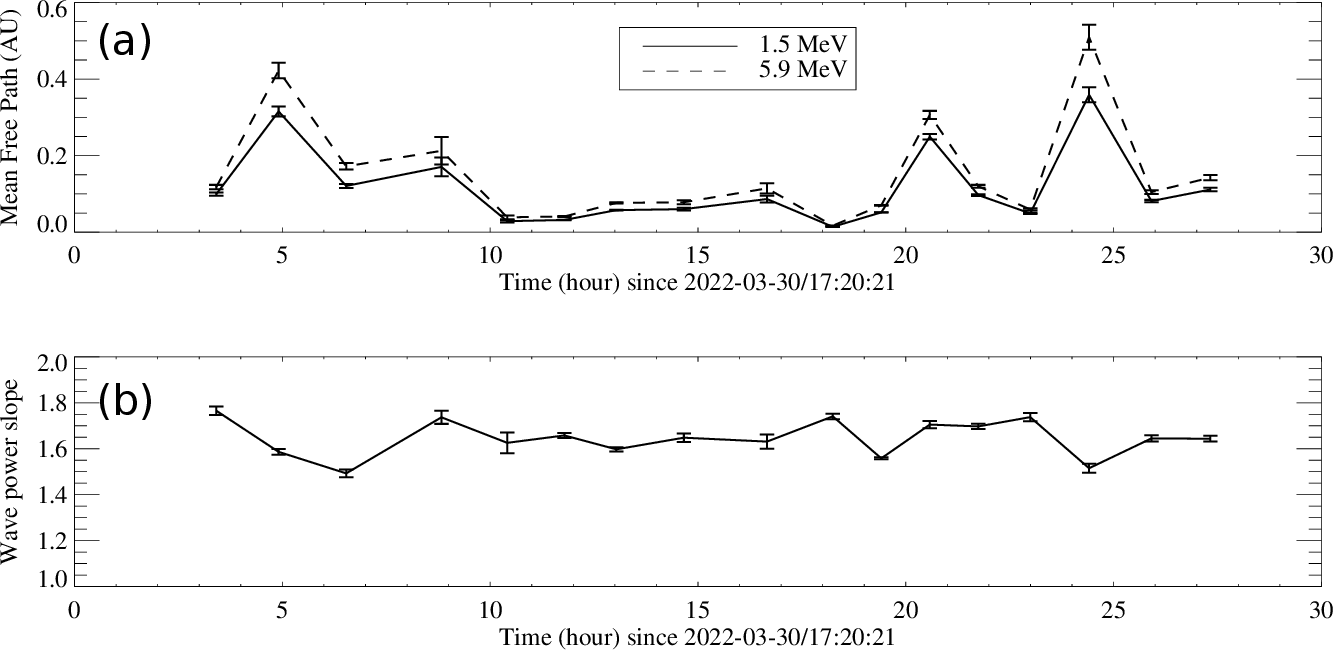}
\caption{Time profiles of (a) the parallel mean free path and (b) the power-law index obtained from the magnetic field analysis \reviser{using the STEREO-A data}. The error bars \reviseb{indicate the 95\% confidence interval estimated from 50 samples}.}
\label{fig:bpower_summ}
\end{figure}

To examine the effect of magnetic field fluctuations on particle transport, we estimate the mean free path \reviser{using the magnetic field data at the STEREO-A location and} the quasi-linear theory \citep{1966ApJ...146..480J}.
The observation interval is manually segmented into multiple time windows based on the occurrence of large magnetic field jumps, as indicated by vertical dashed lines in Figure \ref{fig:sta_pla_mag}.
For each segment, we calculate the mean plasma density $n_0$, velocity $V_0$, and magnetic field vector $\vect{B}_0$.
The transverse magnetic field fluctuations with 1 second cadence, $\delta \vect{B} = \vect{B}-\vect{B}_0$, are decomposed into R- and L-mode circular polarized waves, and then are Fourier transformed using \reviseb{68-minute sequence data (corresponding to 4,096 data points).
Within each segment, 50 ensemble power spectra are randomly selected to evaluate their mean and standard error.}
An example power spectrum density $P(\nu)$ in the frequency range 0.00024-0.5 Hz \reviser{at STEREO-A during 07:00 UT - 09:00 UT} is shown in Figure \ref{fig:bpower_plot}.
The mean proton cyclotron frequency during this period is $\nu_p = \Omega_p / 2\pi =(qB_0)/(2 \pi m_pc) = 0.063 \; {\rm Hz}$.
The dashed lines indicate the resonance frequency for 1.5 MeV and 5.9 MeV protons $(v=17,000 \; \text{and} \; 33,600 \; {\rm km \; s^{-1}})$ given by
\begin{eqnarray}
\nu_R = \nu_p \frac{V_0}{v} = 0.002\left(\frac{B_0}{4.2 \; {\rm nT}}\right) \left(\frac{V_0}{530 \; {\rm km/s}}\right) \left(\frac{v}{1.7 \times 10^4 \; {\rm km/s}}\right)^{-1},\label{eq:7}
\end{eqnarray}
and the effective mean free path parallel to the magnetic field can be estimated in the slab geometry as \citep{1975MNRAS.172..557S}
\begin{eqnarray}
 \lambda_{\parallel} \sim \frac{v}{\Omega_p} \left(\frac{\nu_R P(\nu_R)}{B_0^2}\right)^{-1}.\label{eq:8}
\end{eqnarray}
As the multidimensional nature of magnetic field fluctuations is ignored, this estimate would give the smaller limit \citep{1996JGR...101.2511B}.
To evaluate the power spectrum density at the resonance frequency $P(\nu_R)$, we fit the data between 0.0005 Hz and $\nu_p$ with a power-law function $\propto \nu^{-q}$.
\reviseb{The resulting spectral indices are $q=1.68$ and $1.61$ for R- and L-mode, giving power spectrum density values of 68.9 and 49.8 for 1.5 MeV protons, and 217 and 150 for 5.9 MeV protons, respectively.}
Figure \ref{fig:bpower_summ} shows the time evolution of the estimated parallel mean free path and the spectral index of magnetic field fluctuations \reviser{at 1.0 AU}.
The mean free path decreases over time, approaching 0.1 AU during the SEP decay phase (10-20 hours after 17:20 UT). 
The spectral index remains around 1.5-1.7, which is close to the Kolmogorov turbulence spectrum in the inertial range.

\section{Numerical simulation}\label{sec:numerical-model}
Based on the spacecraft configuration during this event, we expect that the SEPs observed at the BepiColombo location propagated along the nominal Parker spiral field and arrived at the STEREO-A location.
Using an appropriate transport equation, we predict the SEP profile at 1.0 AU from the observed profile at 0.6 AU, and compare it with the observation to assess the model.
For this purpose, we employ the focused transport equation as the governing equation \citep{1995ApJ...442..861R,1998ApJ...509..415L}:
\begin{eqnarray}
\pdif{f}{t} &=& -\pdif{}{z}\left[\mu v + \left\{ 1- \left(\frac{\mu v}{c}\right)^2 \right\}V_{sw}^c\right]f \nonumber \\
&& -\pdif{}{\mu} \left[-\frac{\left(1-\mu^2\right)v}{2} \pdif{\ln |B|}{z} \left\{1+\mu \left(\frac{V_{sw}^c}{v}-\frac{vV_{sw}^c}{c^2}\right)\right\} - \mu \left(1-\mu^2\right) \pdif{V_{sw}^c}{z}\right]f \nonumber \\
&& + \pdif{}{\mu} D_{\mu \mu} \pdif{}{\mu} \left(1-\frac{\mu v V_{sw}^c}{c^2}\right)f,\label{eq:2}
\end{eqnarray}
where $f(z,\mu,t)$ denotes the proton distribution function, $z$ is the spatial coordinate along the mean magnetic field, $\mu$ is the cosine of the pitch angle, $v$ is the proton velocity measured in the local solar wind frame, $c$ is the speed of light, $V_{sw}^c$ is the solar wind velocity in the frame co-rotating with the Sun, and $|B|$ is the magnetic field strength, respectively.
We ignore momentum change by solar wind expansion, as the proton velocity under consideration is much faster than the solar wind velocity.
The magnetic field and solar wind velocity are described by the Parker spiral field model:
\begin{eqnarray}
 B_r &=& B_0 \left(\frac{r_{\rm sun}}{r}\right)^2, \label{eq:3} \\
 B_{\phi} &=& B_r \frac{\Omega_{\rm sun}r}{V_{sw}} \left[\left(\frac{r_{\rm sun}}{r}\right)^2-1\right],\label{eq:4} \\
 V_{sw}^c &=& V_{sw} \frac{|B|}{B_r},\label{eq:5}
\end{eqnarray}
where $r_{\rm sun}$ and $\Omega_{\rm sun}$ are the solar radius and angular velocity, respectively.
We assume a constant solar wind speed of $V_{sw}=600 \; {\rm km \; s^{-1}}$, and the heliocentric distance $r$ is expressed as a function of $z$ using the relation $dr/dz = B_r/|B|$ with $z=0$ at $r=0.6 \; {\rm AU}$.

The deviation from the motion along the mean field is modeled by the pitch-angle diffusion (the third term on the right-hand side of Equation (\ref{eq:2})).
We employ the pitch-angle diffusion coefficient $D_{\mu \mu}$ derived from a dynamical quasi-linear theory \citep{2003ApJ...589.1027D}, which extends the standard theory \citep{1966ApJ...146..480J,1970ApJ...162.1049H} to allow finite scattering through $\mu=0$,
\begin{eqnarray}
D_{\mu \mu} (\mu) &=& \frac{3 v}{2 \left(4-q\right) \left(2-q\right) \lambda_{\parallel}} \left(\mart{\mu^2 + \left(\frac{V_A}{v}\right)^2}\right)^{q-1} \left(1-\mu^2\right),\label{eq:6}\\
\lambda_{\parallel} &=& \xi \frac{v}{\Omega_p},\label{eq:11}
\end{eqnarray}
where $q$ (with $1<q<2$) corresponds to the spectral index of the power spectrum density of magnetic field fluctuations, and $\lambda_{\parallel}$ denotes the mean free path parallel to the magnetic field.
Although both parameters have been estimated from the magnetic field observation at 1.0 AU in \S \ref{sec:solar-wind},  we treat them as free parameters in the numerical simulation.
In Equation (\ref{eq:11}), we express the mean free path proportional to the local cyclotron radius, assuming a spatially uniform factor $\xi$.
We adopt an {\Alfven} speed of $V_A=100 \; {\rm km \; s^{-1}} (\ll v)$, and its appearance in Equation (\ref{eq:6}) reflects the dynamic nature of magnetic turbulence \citep{1994ApJ...420..294B}.


We independently calculate 1.5-5.9 MeV and 5.9-9.1 MeV proton fluxes using the BERM observations at 0.6 AU, $F_{\text{BERM}}$, as the inner boundary condition,
\begin{eqnarray}
f(0,\mu,t) = \frac{F_{\rm BERM}}{v |B|} C \mu^\alpha, \;\;\; {\rm for \;} \mu>0,\label{eq:9}
\end{eqnarray}
where $C$ is a normalization constant, and $v|B|$ serves as a conversion factor from the observation unit (particles cm$^{-2}$ s$^{-1}$) to the simulation unit (particles cm$^{-1}$) by assuming the cross-sectional area of a given flux tube is inversely proportional to the magnetic field strength.
We confirm that the index $\alpha$ has small impact on simulation results discussed below, and therefore use $\alpha=2$.
The simulation domain $z \in [0,2.0\text{AU}]$ (corresponding to $r \in [0.6\text{AU},2.0\text{AU}]$) and $\mu \in [-1,1]$ is discretized into 256 and 32 grid points, respectively.
The resulting proton fluxes at 1.0 AU are then compared with the observed 1.5-5.9 MeV and 5.9-9.1 MeV fluxes at the STEREO-A location, which are estimated by fitting a power-law function to the LET energy channels.

\subsection{Data assimilation}\label{sec:data-assimilation}
To infer the model parameters $(\xi,q)$ that well reproduce the observation at 1.0 AU, we apply a data assimilation technique.
Specifically, we formulate the problem in terms of a state-space model to facilitate the data assimilation using a particle filter \citep{gordon1993novel,kitagawa1996monte}. 
The state vector at time $t$ for an $i$-th ensemble member, denoted as $\vect{x}_i(t)$ for $i=1,\dots,N$, consists of the proton distribution function $f_i(z,\mu,t)$ and the model parameters $\vect{X}_i(t)=(\xi_i(t),q_i(t))$, which determine the pitch-angle diffusion coefficient.
The evolution of the state is governed by the focused transport equation and stochastic updates of the model parameters.
The observation vector at time $t$ corresponds to the proton flux at 1.0 AU taken by LET.
The system noise is introduced by randomly sampling the model parameters at each assimilation step, while the observation noise is assumed to follow a Poisson distribution.
This state-space formulation provides \reviseb{the} basis for sequentially estimating the model parameters through the data assimilation.

At the time $t=t_0$, the model parameters for each ensemble member are sampled from a uniform random distribution,
\begin{eqnarray}
\left\{
\begin{array}{l}
\xi_i \in [5,15000],\\
q_i \in [1.1,1.9],\\
\end{array}
\right.\label{eq:10}
\end{eqnarray}
which serves as a representation of the system noise.
Meanwhile, the same inner boundary condition of Equation (\ref{eq:9}) is applied for all ensemble members.
The distribution function $f_i$ is advanced in time using Equation (\ref{eq:2}) until $t_1 = t_0+\Delta t$, obtaining the predictive distribution of $\vect{x}_i(t_1)$.
The distribution function at 1.0 AU is integrated in phase space to calculate the likelihood $\Lambda_{i}$ \reviseb{that is modeled as a weighted Gaussian function within the assimilation time interval $\Delta  t$} based on the LET data $F_{\text{LET}}$,
\begin{eqnarray}
F_{i} &=& v|B| \inty{-1}{1} f_i d\mu,\label{eq:13}\\
\Lambda_{i} &=& \exp\left[-\frac{1}{2\Delta t} \inty{t_0}{t_1}w(t) \left(\frac{F_{\text{LET}}-F_{i}}{\Delta F_{\text{LET}}}\right)^2 dt\right].\label{eq:14}
\end{eqnarray}
where $w(t) \in [t_0,t_1]$ is a weighting function and $\Delta F_{\text{LET}}$ is the observation noise.
In accordance with the methodology of the particle filter, ensemble members with higher likelihoods are preferentially duplicated, while those with lower likelihoods are dismissed.
This sampling process approximates the posterior probability distribution of $\vect{x}_i(t_1)$.
Subsequently, the model parameters for each ensemble member are sampled in the same manner \reviseb{as those} in Equation (\ref{eq:10}), and the procedure is repeated for the next time step, $t_0 \leftarrow t_1, t_1 \leftarrow t_1+\Delta t$.
This procedure enables the one-step-ahead prediction of the distribution function and the model parameters, forming the basis for sequential data assimilation throughout the simulation period.

Key parameters in this data assimilation are the number of ensemble members $N$ (with larger values generally providing better approximation), the assimilation time interval $\Delta t$, and the weighting function $w(t)$.
To ensure that different sets of the model parameters result in sufficiently distinct distribution function at 1.0 AU, we set the assimilation interval longer than the typical travel time between 0.6 and 1.0 AU (approximately one hour), and the weighting function is an increasing function of $t \in [t_0,t_1]$. 
Accordingly, we use $N=10000$, $\Delta t = 3.8$ hours, and $w(t)=1/[1+(t_1-t)/\Delta t_{\text{LET}}]$ where $\Delta t_{\text{LET}} = 60$ seconds is the time cadence of LET.
The performance of the data assimilation is validated through twin experiments in Appendix, where synthetic observations are generated using known model parameters and then are assimilated to test the prediction of those parameters.

\begin{figure}
\centering
 \includegraphics[clip,angle=0,scale=0.35]{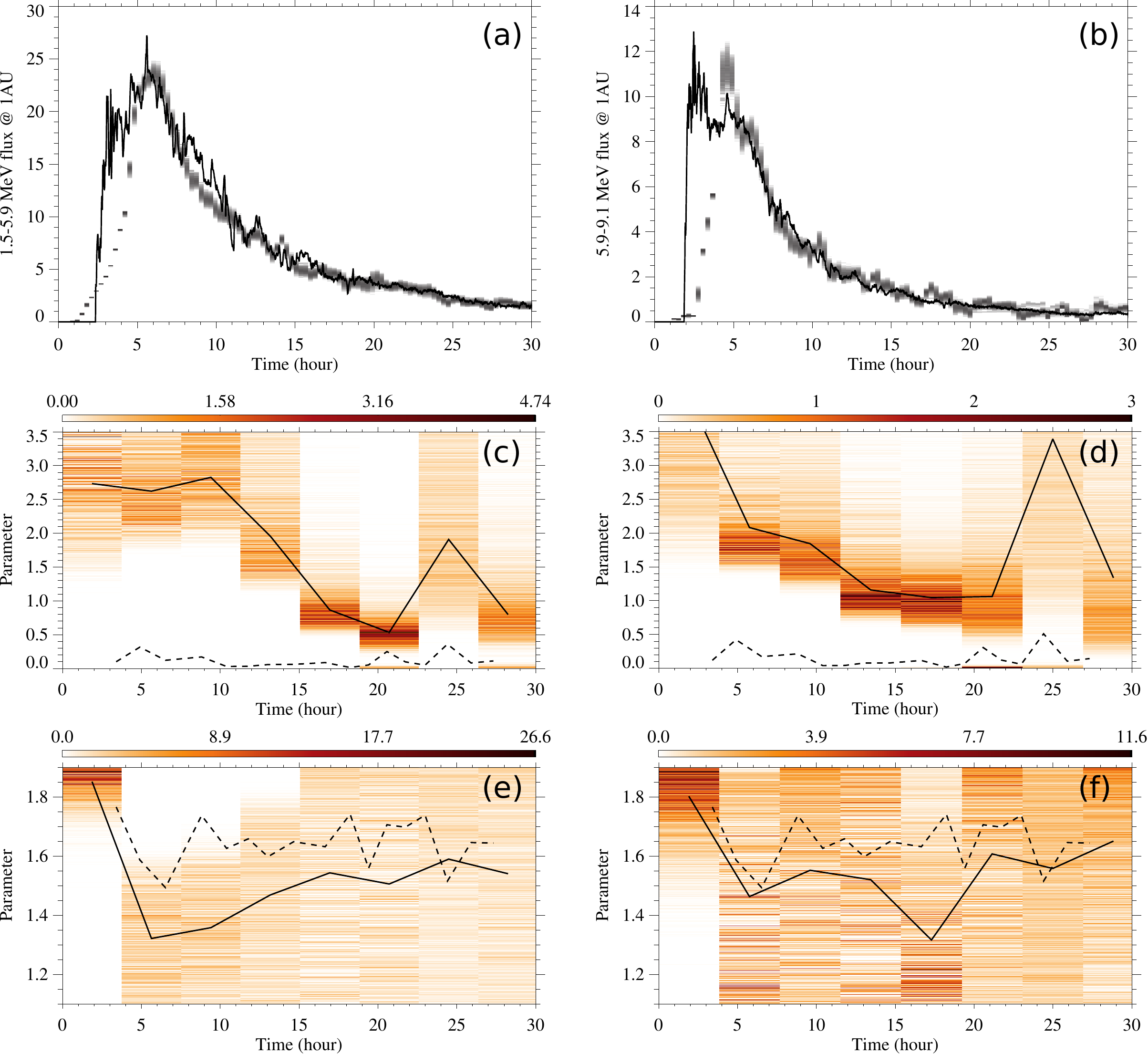}
\caption{Data assimilation into the focused transport simulation. (a,b) Posterior probability distribution of 1.5-5.9 MeV and 5.9-9.1 MeV proton fluxes at 1.0 AU. The black lines represent the STEREO-A/LET observation data. (c,d) Posterior probability distribution of the mean free path $\lambda_{\parallel}$ (AU) for 1.5 MeV and 5.9 MeV protons. (e,f) Posterior probability distribution of the spectral index $q$ in Equation (\ref{eq:6}). In panels (c-f), the black solid lines represent the median values, while the dashed lines are obtained from the magnetic field analysis in Figure \ref{fig:bpower_summ}.} 
\label{fig:da_plt_all}
\end{figure}

Figure \ref{fig:da_plt_all}(a) and (b) compare the time profiles of 1.5-5.9 MeV and 5.9-9.1 MeV proton fluxes at 1.0 AU, obtained from the observation (black lines) and the simulations (\reviser{gray shading}).
The simulations fail to reproduce the preceding component seen in LET ($t<4$ hours), as is expected by the fact that a counterpart of the preceding component is not evident in BERM (Figure \ref{fig:berm_let_prof}).
After that ($t>4$ hours), the simulations successfully reproduce the observed profiles, indicating that BERM and LET observed the same population during this period.
Figure \ref{fig:da_plt_all}(c) and (d) present the time evolution of the probability distribution of the mean free path $\lambda_{\parallel}$ for 1.5 MeV and 5.9 MeV protons.
The median values (black solid lines) decrease over time, approaching approximately 0.5 AU and 1.0 AU for 1.5 MeV and 5.9 MeV protons during the decay phase ($t=12 - 23$ hours).
Subsequently, both channels show an increase in $\lambda_{\parallel}$ at $t=25$ \reviseb{hours}, followed by a return to small values.
The decrease in $\lambda_{\parallel}$ during the decay phase reflects enhanced pitch-angle scattering, suggesting the transition from ballistic to diffusive propagation over time.
The energy dependence of $\lambda_{\parallel}$ is not clearly established: for $t>14$ hours, it is longer for 5.9 MeV protons, while the opposite trend is seen for $t<14$ hours.
Figure \ref{fig:da_plt_all}(e) and (f) show the probability distribution of the spectral index $q$.
Except during the first period, they are broadly distributed without a clear systematic trend, indicating that the integral flux used for the data assimilation (Eqs. (\ref{eq:13})-~(\ref{eq:14})) is insensitive to the choice of the spectral index.


\section{Summary and discussion}\label{sec:dis}
The SEP event on March 30, 2022, was observed simultaneously by the magnetically well-connected BepiColombo and STEREO-A spacecraft at heliocentric distances of 0.6 AU and 1.0 AU, providing a valuable opportunity to investigate particle transport processes in the inner heliosphere.
While only the long-lasting component was observed at the BepiColombo location, the SEP profile exhibited two distinct components at the STEREO-A location: the highly-fluctuating preceding component and the long-lasting component.
The velocity dispersion was observed by both spacecraft during the rise phase.
The inferred path length and the release time from the velocity dispersion analysis are 0.56 AU and 19:42 UT from BepiColombo, and 0.92 AU and 18:12 UT from STEREO-A (for the preceding component), respectively.
The CME heights at these periods are obtained from the SOHO/LASCO CME Catalog \citep{2009EM&P..104..295G}, 0.019AU at 18:12 UT and 0.041 AU at 19:42 UT.
The total path length is 0.6 AU from the Sun to BepiColombo, indicating the CME as a possible source of the long-lasting component.

In contrast, the origin of the preceding component observed only by STEREO-A remains ambiguous.
\reviser{Its onset time is much earlier than that observed by BepiColombo, even though STEREO-A was expected to be magnetically connected to a region slightly farther from the flare site than BepiColombo.}
The path length from the Sun to STEREO-A, inferred from the preceding component, is shorter than 1.0 AU even when accounting for the projection effect in the CME height, suggesting that this component is unlikely to have originated from the CME.
Another possible source would be a halo wave initiated by the flare, which propagates through the solar corona and precedes the CME, as observed by the SDO/AIA and SOHO/LASCO instruments.
However, the propagation speed to meet the SEP observation, \revrr{0.13 AU/(18:12 UT - 17:20 UT) = $6200 \; {\rm km \; s^{-1}}$}, exceeds typical fast magnetosonic wave speeds in the solar corona, making this interpretation uncertain.


The absence of the preceding component at the BepiColombo location may be attributed to spacecraft and/or solar wind conditions.
Since the field of view of BERM is limited and directed away from the Sun, it may not detect highly collimated particles streaming from the Sun.
The preceding component during the rise phase may correspond to such particles.
This effect would diminish as the pitch-angle distribution is relaxed over time.
Alternatively, the magnetic connectivity between BepiColombo and STEREO-A may have been weak or less effective during the rise phase, owing to differences in solar wind speed.
\reviser{As shown in Figure \ref{fig:solarmach}, the longitudinal separation of their magnetic footpoint on the solar surface is expected to be larger during the rise phase than during the decay phase.
The magnetic connectivity between the two spacecraft likely improved over time as the solar wind speed changed, which could account for the initial discrepancy and the subsequent correspondence of the SEP time profiles during the rise and decay phases.
Meanwhile, a similar decreasing time profile of the total magnetic field strength during the rise phase was observed at both locations (Figure \ref{fig:sta_pla_mag}(f)), which appears to place BepiColombo and STEREO-A on a similar Parker spiral.
Therefore, the scenario is not conclusive.
Local magnetic field structures may also play a role in the absence of such components, which remains an issue for future investigation.}

The SEPs during the decay phase are characterized by long-lasting time profiles, with longer time scales observed at the STEREO-A location compared to the BepiColombo location, indicating the deviation from ballistic propagation.
We model this result using the focused transport equation including the pitch-angle diffusion.
To determine the model parameters used in the pitch-angle diffusion coefficient, for the first time, we assimilate the multi-spacecraft observation data into the focused transport simulations, enabling us to estimate the mean free path as a time series.
From the peak to the decay phase, the simulations reproduce the observed profiles, indicating that BepiColombo and STEREO-A observed the same population during this period.
The inferred mean free path decreases to approximately 0.5 AU for 1.5 MeV protons and 1.0 AU for 5.9 MeV protons, suggesting an increasing influence of scattering on particle transport.
To validate these results, we independently estimate the mean free path from magnetic field fluctuations observed at the STEREO-A location.
The temporal variation of this estimate qualitatively agrees with the trend derived from the SEP data assimilation, supporting the connection between the particle transport and the local field fluctuations.
However, the mean free path estimated from the magnetic field fluctuations is 5-10 times shorter than that from the data assimilation: it is approximately 0.3 AU during the rise phase, and 0.1 AU during the decay phase while the data assimilation indicates the magnitude longer than 0.5 AU.
Considering that the velocity dispersion is observed during the rise phase, the mean free path during this period should be much longer than 1.0 AU.
Therefore, it is likely that the mean free path from the magnetic field fluctuations is underestimated.

The discrepancy in the magnitude of the mean free path estimated from SEP observations and magnetic field fluctuations has been investigated thus far.
\cite{1994ApJ...420..294B} suggested that magnetic field fluctuations include a substantial fraction of perpendicular components, which contribute little to particle scattering.
As a result, the mean free path may be underestimated when assuming a purely slab geometry.
\cite{1996JGR...101.2511B} presented that about 20\% of the total magnetic field fluctuations correspond to parallel components that are effective in scattering particles.
Based on this assumption, the mean free path estimated from the standard quasi-linear theory can be scaled up by a factor of five, bringing it closer to the data assimilation estimates.

By incorporating the power-law model for the power spectrum density into the quasi-linear theory, the spectral index $q$ appears in the formulation of the pitch-angle diffusion coefficient, and can be compared between the magnetic field observations and the data assimilation results.
While the observed magnetic field fluctuations yield an index around 1.6, the data assimilation results cannot provide a conclusive estimate.
The index governs the rigidity (momentum) dependence of the mean free path through the relation $\lambda_{\parallel} \propto p^{2-q}$ under the standard theory.
However, the data assimilation results do not show clear trend in the rigidity dependence, preventing a conclusive assessment of its consistency with the index.
The data assimilation employed in this study relies solely on the integral flux at 1.0 AU.
Incorporating pitch-angle distribution measurements in future assimilation efforts would better constrain the pitch-angle diffusion coefficient and validate the relation between $\lambda_{\parallel}$ and $q$.

The present study only considers the diffusive transport parallel to the magnetic field lines, and ignore cross-field diffusion arising from particle scattering and drifting, and magnetic field line meandering.
Cross-field diffusion is typically included in models of galactic cosmic ray transport into the heliosphere.
The amount of cross-field diffusion is discussed in terms of the ratio of perpendicular to parallel mean free paths (or diffusion coefficients), $\lambda_{\perp}/\lambda_{\parallel}$, with reported values for SEPs spanning a wide range from $10^{-4}$ to unity \citep[][and references therein]{2014JGRA..119.6074D}.
In this event, we constrain the perpendicular mean free path from the fact that significant enhancement of protons with energies above 3.4 MeV was not observed at Earth until 05:00 UT on March 31, while they were detected at 18:00 UT on March 30 by STEREO-A located $33^{\circ}$ east of Earth.
Therefore, the azimuthal diffusion coefficient and the perpendicular mean free path may be limited by
\begin{eqnarray}
\kappa_{\phi \phi} &\lsim& \frac{(33 \pi/180)^2}{11\times3600} = 8.4\times 10^{-6} \; s^{-1},\label{eq:16}\\
A^2 \kappa_{\phi \phi} &=& \frac{v}{3} \left(\lambda_{\parallel} \sin^2 \Psi + \lambda_{\perp} \cos^2 \Psi\right)\nonumber \\
&\Rightarrow& \lambda_{\perp} \lsim \frac{ 3 A^2 \kappa_{\phi \phi}}{v \cos^2 \Psi},\label{eq:19}  
\end{eqnarray}
where $\Psi=\tan^{-1}(B_{\phi}/B_r)$ and $A = 1$ AU.
For particle and solar wind speeds of $33,600 \; {\rm km \; s^{-1}}$ (5.9 MeV protons) and $600 \; {\rm km \; s^{-1}}$ $(\cos \Psi = 0.81)$, 
 we obtain $\lambda_{\perp} \lsim 0.17 \; {\rm AU}$, which is shorter than the parallel mean free path inferred from the SEP data assimilation.
This upper limit on the perpendicular mean free path is consistent with the minimal role of cross-field diffusion, supporting the use of the one-dimensional transport model along the Parker spiral field in the present study.
Multi-spacecraft observations at different heliocentric longitudes are valuable for discussing the amount of cross-field diffusion and its relationship to parallel diffusion, especially in impulsive events from a narrow area \citep{2014JGRA..119.6074D,2016ApJ...826..134D}.

By comparing with Equations (\ref{eq:8}) and (\ref{eq:11}), our model assumes that the shape of the power spectrum is the same at different heliocentric distances, although this has not been necessarily validated by observations.
Under this assumption, the parallel mean free path is inversely proportional to the local magnetic field strength, $\lambda_{\parallel} \propto |B|^{-1} \propto r^{b}$ where $b = 2$ for $r \ll A$ and $b = 1$ for $r \gg A$.
The radial dependence of the mean free path is a significant factor in modeling particle transport.
Earlier observational estimates in the outer heliosphere, summarized by \cite{1982RvGSP..20..335P}, suggest $b=0.9$ to $1.8$ over 1.0-5.0 AU.
More recently, \cite{2024ApJ...965...61C} analyzed magnetic field data taken by Parker Solar Probe to determine the radial dependence in the inner heliosphere, yielding $b=1.17$ over 0.1-0.8 AU.
An integrated study combining multi-spacecraft observations of SEPs and magnetic fields in both the inner and outer heliosphere, focused transport simulations with realistic observation-based models, and their assimilation is planned in a future work for comprehensive understanding of the transport process of SEPs.

\section*{Declarations}

\section*{Availability of data and materials}
\reviseb{The current study analyzed the STEREO/LET Level 1 combined data taken from the data center at the California Institute of Technology, and the PLASTIC and IMPACT Level 2 data taken from the data server at University of California, Los Angeles.
The GOES-16 SGPX data is downloaded through the SPEDAS data analysis tool.}
The datasets used and/or analyzed during the current study are available from the corresponding author upon reasonable request.

\section*{Competing interests}
The authors declare that they have no competing interests.

\section*{Funding}
TM is supported by JSPS KAKENHI Grant Number JP25H00625.
KI is supported by JSPS KAKENHI Grant Numbers 24H00022 and 21H04517.

\section*{Authors' contributions}
TM and YM designed the concept and methodology of the study. 
GM and MP provided the MPO/BERM data. 
\reviser{DS, AM, and WB provided the Mio/MGF data, with cleaning and calibration by DF.}
GM, KI, and SI supported the data analysis and numerical simulations conducted by TM and YM. 
TM wrote the original draft, and other authors reviewed and edited it.

\acknowledgments{
We thank the anonymous reviewers for their careful reviewing of the manuscript and for their constructive comments, which greatly improved the paper.
We thank to BepiColombo, STEREO, and GOES teams for providing spacecraft data, and D. Shiota for his valuable comments on the CME observation.
This work was carried out by the joint research program of Institute for Space-Earth Environmental Research, Nagoya University.
}

\appendix{Twin experiments}\label{sec:twin-experiments}
To assess the capability of the data assimilation into the focused transport simulation to infer the model parameters, we perform twin experiments.
Firstly, three sets of synthetic observation data for 1.5-5.9 MeV protons at 1.0 AU are generated using the simulation with $q=1.6$, Equation (\ref{eq:9}) as the inner boundary condition, and \reviseb{the mean free path models,}
\begin{eqnarray}
\lambda_{\parallel} =
\left\{
\begin{array}{l}
1.18 \exp[-t_h/13.9], \\
0.83, \\
1.18 \exp[t_h/27.8],
\end{array}
\right.\label{eq:15}
\end{eqnarray}
where \reviseb{$\lambda_{\parallel}$ is measured in AU} and $t_h$ is the time measured in hour.
Subsequently, these synthetic data are assimilated into the simulations following the same procedure in \S~\ref{sec:data-assimilation}, using \reviseb{an ensemble size of $N=1000$}.

\begin{figure}
 \centering
\includegraphics[clip,angle=0,scale=0.35]{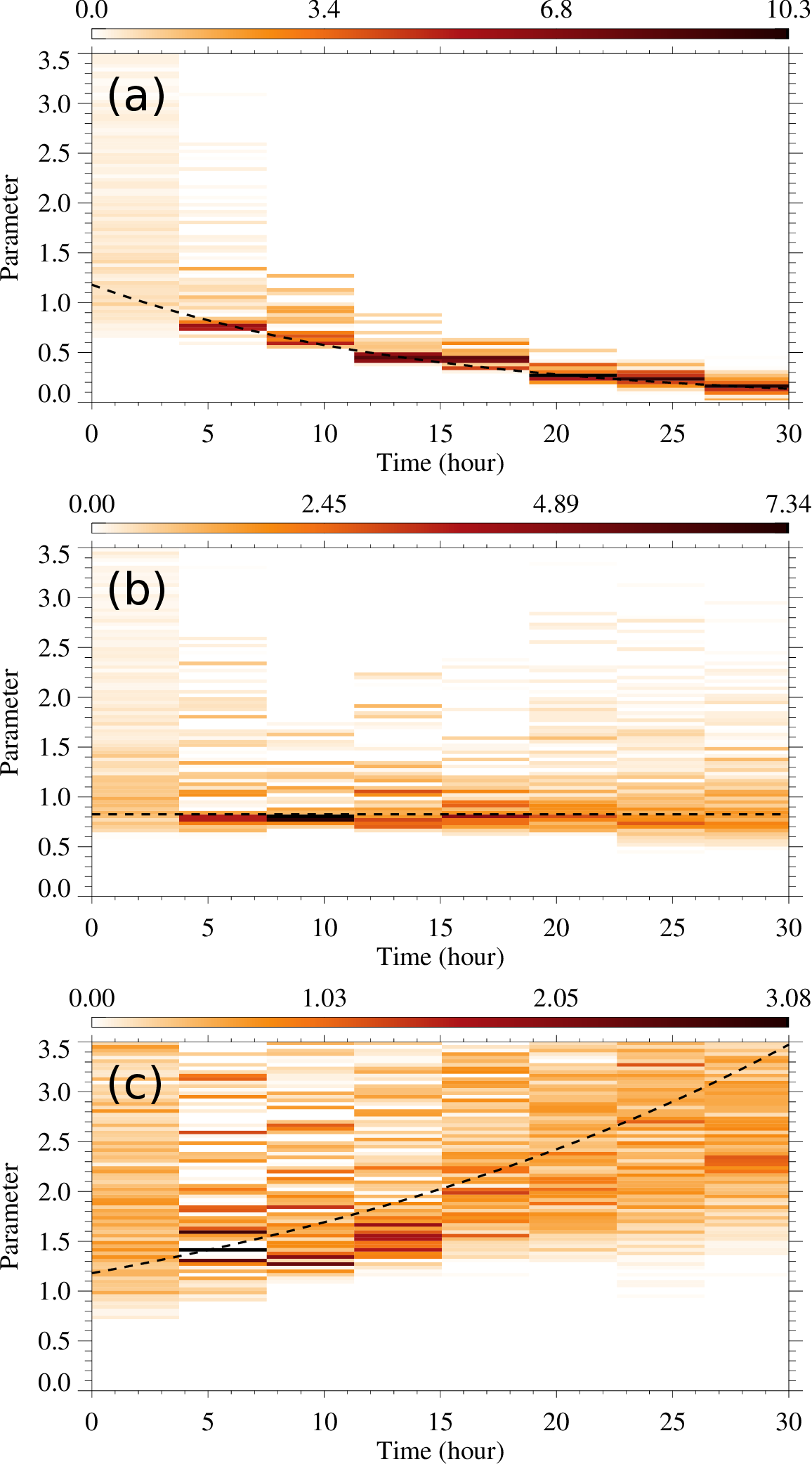}
\caption{Posterior probability distribution of the mean free path $\lambda_{\parallel}$ (AU) for 1.5 MeV protons in the twin experiments. (a,b,c) Results for decreasing, constant, and increasing mean free path models. The dashed lines represent the answer.}
\label{fig:da_test}
\end{figure}

Figure \ref{fig:da_test} shows the probability distributions of the inferred mean free path for three test cases.
The results demonstrate that the data assimilation successfully predicts both the decreasing and constant mean free path models (panels (a) and (b)), where the magnitude is shorter than 1.0 AU.
In contrast, the prediction for the increasing mean free path model (c) is less accurate.
This is acceptable, as the propagation in this case is approximately scatter-free, making the solution insensitive to variations in the mean free path exceeding 1.0 AU.

\bibliographystyle{plainnat}
\bibliography{ref}

@ARTICLE{1999SSRv...90..413R,
   author = {{Reames}, D.~V.},
    title = "{Particle acceleration at the Sun and in the heliosphere}",
  journal = {Space Science Reviews},
     year = 1999,
    month = oct,
   volume = 90,
    pages = {413-491},
      doi = {10.1023/A:1005105831781},
   adsurl = {http://ads.nao.ac.jp/abs/1999SSRv...90..413R}
}

@ARTICLE{2023FrASS...958810G,
       author = {{Gieseler}, Jan and {Dresing}, Nina and {Palmroos}, Christian and {Freiherr von Forstner}, Johan L. and {Price}, Daniel J. and {Vainio}, Rami and {Kouloumvakos}, Athanasios and {Rodr{\'\i}guez-Garc{\'\i}a}, Laura and {Trotta}, Domenico and {G{\'e}not}, Vincent and {Masson}, Arnaud and {Roth}, Markus and {Veronig}, Astrid},
        title = "{Solar-MACH: An open-source tool to analyze solar magnetic connection configurations}",
      journal = {Frontiers in Astronomy and Space Sciences},
         year = 2023,
        month = feb,
       volume = {9},
          eid = {384},
        pages = {384},
          doi = {10.3389/fspas.2022.1058810},
archivePrefix = {arXiv},
       eprint = {2210.00819},
 primaryClass = {astro-ph.SR},
       adsurl = {https://ui.adsabs.harvard.edu/abs/2023FrASS...958810G}
}

@ARTICLE{2022SSRv..218...54P,
       author = {{Pinto}, Marco and {Sanchez-Cano}, Beatriz and {Moissl}, Richard and {Benkhoff}, Johannes and {Cardoso}, Carlota and {Gon{\c{c}}alves}, Patr{\'\i}cia and {Assis}, Pedro and {Vainio}, Rami and {Oleynik}, Philipp and {Lehtolainen}, Arto and {Grande}, Manuel and {Marques}, Arlindo},
        title = "{The BepiColombo Environment Radiation Monitor, BERM}",
      journal = {\ssr},
         year = 2022,
        month = oct,
       volume = {218},
       number = {7},
          eid = {54},
        pages = {54},
          doi = {10.1007/s11214-022-00922-2},
       adsurl = {https://ui.adsabs.harvard.edu/abs/2022SSRv..218...54P}
}

@ARTICLE{2021SSRv..217...90B,
       author = {{Benkhoff}, J. and {Murakami}, G. and {Baumjohann}, W. and {Besse}, S. and {Bunce}, E. and {Casale}, M. and {Cremonese}, G. and {Glassmeier}, K. -H. and {Hayakawa}, H. and {Heyner}, D. and {Hiesinger}, H. and {Huovelin}, J. and {Hussmann}, H. and {Iafolla}, V. and {Iess}, L. and {Kasaba}, Y. and {Kobayashi}, M. and {Milillo}, A. and {Mitrofanov}, I.~G. and {Montagnon}, E. and {Novara}, M. and {Orsini}, S. and {Quemerais}, E. and {Reininghaus}, U. and {Saito}, Y. and {Santoli}, F. and {Stramaccioni}, D. and {Sutherland}, O. and {Thomas}, N. and {Yoshikawa}, I. and {Zender}, J.},
        title = "{BepiColombo - Mission Overview and Science Goals}",
      journal = {\ssr},
         year = 2021,
        month = dec,
       volume = {217},
       number = {8},
          eid = {90},
        pages = {90},
          doi = {10.1007/s11214-021-00861-4},
       adsurl = {https://ui.adsabs.harvard.edu/abs/2021SSRv..217...90B}
}

@ARTICLE{2020SSRv..216..113M,
       author = {{Murakami}, Go and {Hayakawa}, Hajime and {Ogawa}, Hiroyuki and {Matsuda}, Shoya and {Seki}, Taeko and {Kasaba}, Yasumasa and {Saito}, Yoshifumi and {Yoshikawa}, Ichiro and {Kobayashi}, Masanori and {Baumjohann}, Wolfgang and {Matsuoka}, Ayako and {Kojima}, Hirotsugu and {Yagitani}, Satoshi and {Moncuquet}, Michel and {Wahlund}, Jan-Erik and {Delcourt}, Dominique and {Hirahara}, Masafumi and {Barabash}, Stas and {Korablev}, Oleg and {Fujimoto}, Masaki},
        title = "{Mio{\textemdash}First Comprehensive Exploration of Mercury's Space Environment: Mission Overview}",
      journal = {\ssr},
         year = 2020,
        month = oct,
       volume = {216},
       number = {7},
          eid = {113},
        pages = {113},
          doi = {10.1007/s11214-020-00733-3},
       adsurl = {https://ui.adsabs.harvard.edu/abs/2020SSRv..216..113M}
}

@ARTICLE{2008SSRv..136..285M,
       author = {{Mewaldt}, R.~A. and {Cohen}, C.~M.~S. and {Cook}, W.~R. and {Cummings}, A.~C. and {Davis}, A.~J. and {Geier}, S. and {Kecman}, B. and {Klemic}, J. and {Labrador}, A.~W. and {Leske}, R.~A. and {Miyasaka}, H. and {Nguyen}, V. and {Ogliore}, R.~C. and {Stone}, E.~C. and {Radocinski}, R.~G. and {Wiedenbeck}, M.~E. and {Hawk}, J. and {Shuman}, S. and {von Rosenvinge}, T.~T. and {Wortman}, K.},
        title = "{The Low-Energy Telescope (LET) and SEP Central Electronics for the STEREO Mission}",
      journal = {\ssr},
         year = 2008,
        month = apr,
       volume = {136},
       number = {1-4},
        pages = {285-362},
          doi = {10.1007/s11214-007-9288-x},
       adsurl = {https://ui.adsabs.harvard.edu/abs/2008SSRv..136..285M}
}

@ARTICLE{1995ApJ...442..861R,
       author = {{Ruffolo}, D.},
        title = "{Effect of Adiabatic Deceleration on the Focused Transport of Solar Cosmic Rays}",
      journal = {\apj},
         year = 1995,
        month = apr,
       volume = {442},
        pages = {861},
          doi = {10.1086/175489},
archivePrefix = {arXiv},
       eprint = {astro-ph/9408056},
 primaryClass = {astro-ph},
       adsurl = {https://ui.adsabs.harvard.edu/abs/1995ApJ...442..861R}
}

@ARTICLE{1998ApJ...509..415L,
       author = {{Lario}, D. and {Sanahuja}, B. and {Heras}, A.~M.},
        title = "{Energetic Particle Events: Efficiency of Interplanetary Shocks as 50 keV < E < 100 MeV Proton Accelerators}",
      journal = {\apj},
         year = 1998,
        month = dec,
       volume = {509},
       number = {1},
        pages = {415-434},
          doi = {10.1086/306461},
       adsurl = {https://ui.adsabs.harvard.edu/abs/1998ApJ...509..415L}
}

@ARTICLE{1975MNRAS.172..557S,
       author = {{Skilling}, J.},
        title = "{Cosmic ray streaming - I. Effect of Alfv{\'e}n waves on particles.}",
      journal = {\mnras},
         year = 1975,
        month = sep,
       volume = {172},
        pages = {557-566},
          doi = {10.1093/mnras/172.3.557},
       adsurl = {https://ui.adsabs.harvard.edu/abs/1975MNRAS.172..557S}
}

@ARTICLE{2016LRSP...13....3D,
       author = {{Desai}, Mihir and {Giacalone}, Joe},
        title = "{Large gradual solar energetic particle events}",
      journal = {Living Reviews in Solar Physics},
         year = 2016,
        month = dec,
       volume = {13},
       number = {1},
          eid = {3},
        pages = {3},
          doi = {10.1007/s41116-016-0002-5},
       adsurl = {https://ui.adsabs.harvard.edu/abs/2016LRSP...13....3D}
}

@ARTICLE{1982RvGSP..20..335P,
       author = {{Palmer}, I.~D.},
        title = "{Transport coefficients of low-energy cosmic rays in interplanetary space}",
      journal = {Reviews of Geophysics and Space Physics},
         year = 1982,
        month = may,
       volume = {20},
        pages = {335-351},
          doi = {10.1029/RG020i002p00335},
       adsurl = {https://ui.adsabs.harvard.edu/abs/1982RvGSP..20..335P}
}

@ARTICLE{2024ApJ...971..105L,
       author = {{Lang}, J.~T. and {Strauss}, R.~D. and {Engelbrecht}, N.~E. and {van den Berg}, J.~P. and {Dresing}, N. and {Ruffolo}, D. and {Bandyopadhyay}, R.},
        title = "{A Detailed Survey of the Parallel Mean Free Path of Solar Energetic Particle Protons and Electrons}",
      journal = {\apj},
         year = 2024,
        month = aug,
       volume = {971},
       number = {1},
          eid = {105},
        pages = {105},
          doi = {10.3847/1538-4357/ad55c3},
archivePrefix = {arXiv},
       eprint = {2406.05765},
 primaryClass = {astro-ph.SR},
       adsurl = {https://ui.adsabs.harvard.edu/abs/2024ApJ...971..105L}
}

@ARTICLE{2024JSWSC..14....3P,
       author = {{Palmerio}, Erika and {Luhmann}, Janet G. and {Mays}, M. Leila and {Caplan}, Ronald M. and {Lario}, David and {Richardson}, Ian G. and {Whitman}, Kathryn and {Lee}, Christina O. and {S{\'a}nchez-Cano}, Beatriz and {Wijsen}, Nicolas and {Li}, Yan and {Cardoso}, Carlota and {Pinto}, Marco and {Heyner}, Daniel and {Schmid}, Daniel and {Auster}, Hans-Ulrich and {Fischer}, David},
        title = "{Improved modelling of SEP event onset within the WSA-Enlil-SEPMOD framework}",
      journal = {Journal of Space Weather and Space Climate},
         year = 2024,
        month = feb,
       volume = {14},
          eid = {3},
        pages = {3},
          doi = {10.1051/swsc/2024001},
archivePrefix = {arXiv},
       eprint = {2401.05309},
 primaryClass = {astro-ph.SR},
       adsurl = {https://ui.adsabs.harvard.edu/abs/2024JSWSC..14....3P}
}

@ARTICLE{2000ApJ...537.1073D,
       author = {{Dr{\"o}ge}, Wolfgang},
        title = "{The Rigidity Dependence of Solar Particle Scattering Mean Free Paths}",
      journal = {\apj},
         year = 2000,
        month = jul,
       volume = {537},
       number = {2},
        pages = {1073-1079},
          doi = {10.1086/309080},
       adsurl = {https://ui.adsabs.harvard.edu/abs/2000ApJ...537.1073D}
}

@ARTICLE{2003ApJ...589.1027D,
       author = {{Dr{\"o}ge}, Wolfgang},
        title = "{Solar Particle Transport in a Dynamical Quasi-linear Theory}",
      journal = {\apj},
         year = 2003,
        month = jun,
       volume = {589},
       number = {2},
        pages = {1027-1039},
          doi = {10.1086/374812},
       adsurl = {https://ui.adsabs.harvard.edu/abs/2003ApJ...589.1027D}
}

@ARTICLE{2009ApJ...693...69D,
       author = {{Dr{\"o}ge}, W. and {Kartavykh}, Y.~Y.},
        title = "{Testing Transport Theories with Solar Energetic Particles}",
      journal = {\apj},
         year = 2009,
        month = mar,
       volume = {693},
       number = {1},
        pages = {69-74},
          doi = {10.1088/0004-637X/693/1/69},
       adsurl = {https://ui.adsabs.harvard.edu/abs/2009ApJ...693...69D}
}

@ARTICLE{2014JGRA..119.6074D,
       author = {{Dr{\"o}ge}, W. and {Kartavykh}, Y.~Y. and {Dresing}, N. and {Heber}, B. and {Klassen}, A.},
        title = "{Wide longitudinal distribution of interplanetary electrons following the 7 February 2010 solar event: Observations and transport modeling}",
      journal = {Journal of Geophysical Research (Space Physics)},
         year = 2014,
        month = aug,
       volume = {119},
       number = {8},
        pages = {6074-6094},
          doi = {10.1002/2014JA019933},
       adsurl = {https://ui.adsabs.harvard.edu/abs/2014JGRA..119.6074D}
}

@ARTICLE{2016ApJ...826..134D,
       author = {{Dr{\"o}ge}, W. and {Kartavykh}, Y.~Y. and {Dresing}, N. and {Klassen}, A.},
        title = "{Multi-spacecraft Observations and Transport Modeling of Energetic Electrons for a Series of Solar Particle Events in August 2010}",
      journal = {\apj},
         year = 2016,
        month = aug,
       volume = {826},
       number = {2},
          eid = {134},
        pages = {134},
          doi = {10.3847/0004-637X/826/2/134},
       adsurl = {https://ui.adsabs.harvard.edu/abs/2016ApJ...826..134D}
}

@ARTICLE{1994ApJ...420..294B,
       author = {{Bieber}, John W. and {Matthaeus}, William H. and {Smith}, Charles W. and {Wanner}, Wolfgang and {Kallenrode}, May-Britt and {Wibberenz}, Gerd},
        title = "{Proton and Electron Mean Free Paths: The Palmer Consensus Revisited}",
      journal = {\apj},
         year = 1994,
        month = jan,
       volume = {420},
        pages = {294},
          doi = {10.1086/173559},
       adsurl = {https://ui.adsabs.harvard.edu/abs/1994ApJ...420..294B}
}

@ARTICLE{1996JGR...101.2511B,
       author = {{Bieber}, John W. and {Wanner}, Wolfgang and {Matthaeus}, William H.},
        title = "{Dominant two-dimensional solar wind turbulence with implications for cosmic ray transport}",
      journal = {\jgr},
         year = 1996,
        month = feb,
       volume = {101},
       number = {A2},
        pages = {2511-2522},
          doi = {10.1029/95JA02588},
       adsurl = {https://ui.adsabs.harvard.edu/abs/1996JGR...101.2511B}
}

@ARTICLE{1993JGR....98.3513W,
       author = {{Wanner}, Wolfgang and {Wibberenz}, Gerd},
        title = "{A study of the propagation of solar energetic protons in the inner heliosphere}",
      journal = {\jgr},
         year = 1993,
        month = mar,
       volume = {98},
       number = {A3},
        pages = {3513-3528},
          doi = {10.1029/92JA02546},
       adsurl = {https://ui.adsabs.harvard.edu/abs/1993JGR....98.3513W}
}

@ARTICLE{1993AdSpR..13i.359W,
       author = {{Wanner}, W. and {Kallenrode}, M. -B. and {Dr{\"o}ge}, W. and {Wibberenz}, G.},
        title = "{Solar energetic proton mean free paths}",
      journal = {Advances in Space Research},
         year = 1993,
        month = sep,
       volume = {13},
       number = {9},
        pages = {359-362},
          doi = {10.1016/0273-1177(93)90505-6},
       adsurl = {https://ui.adsabs.harvard.edu/abs/1993AdSpR..13i.359W}
}

@ARTICLE{2005ApJ...627..562Q,
       author = {{Qin}, G. and {Zhang}, M. and {Dwyer}, J.~R. and {Rassoul}, H.~K. and {Mason}, G.~M.},
        title = "{The Model Dependence of Solar Energetic Particle Mean Free Paths under Weak Scattering}",
      journal = {\apj},
         year = 2005,
        month = jul,
       volume = {627},
       number = {1},
        pages = {562-566},
          doi = {10.1086/430136},
       adsurl = {https://ui.adsabs.harvard.edu/abs/2005ApJ...627..562Q}
}

@ARTICLE{2006JGRA..111.8101Q,
       author = {{Qin}, G. and {Zhang}, M. and {Dwyer}, J.~R.},
        title = "{Effect of adiabatic cooling on the fitted parallel mean free path of solar energetic particles}",
      journal = {Journal of Geophysical Research (Space Physics)},
         year = 2006,
        month = aug,
       volume = {111},
       number = {A8},
          eid = {A08101},
        pages = {A08101},
          doi = {10.1029/2005JA011512},
       adsurl = {https://ui.adsabs.harvard.edu/abs/2006JGRA..111.8101Q}
}

@ARTICLE{2004A&A...420..343L,
       author = {{Lintunen}, J. and {Vainio}, R.},
        title = "{Solar energetic particle event onset as analyzed from simulated data}",
      journal = {\aap},
         year = 2004,
        month = jun,
       volume = {420},
        pages = {343-350},
          doi = {10.1051/0004-6361:20034247},
       adsurl = {https://ui.adsabs.harvard.edu/abs/2004A&A...420..343L}
}

@ARTICLE{2013JGRA..118..642T,
       author = {{Tautz}, R.~C. and {Shalchi}, A.},
        title = "{Simulated energetic particle transport in the interplanetary space: The Palmer consensus revisited}",
      journal = {Journal of Geophysical Research (Space Physics)},
         year = 2013,
        month = feb,
       volume = {118},
       number = {2},
        pages = {642-647},
          doi = {10.1002/jgra.50155},
archivePrefix = {arXiv},
       eprint = {1301.7162},
 primaryClass = {astro-ph.HE},
       adsurl = {https://ui.adsabs.harvard.edu/abs/2013JGRA..118..642T}
}

@ARTICLE{1966ApJ...146..480J,
       author = {{Jokipii}, J.~R.},
        title = "{Cosmic-Ray Propagation. I. Charged Particles in a Random Magnetic Field}",
      journal = {\apj},
         year = 1966,
        month = nov,
       volume = {146},
        pages = {480},
          doi = {10.1086/148912},
       adsurl = {https://ui.adsabs.harvard.edu/abs/1966ApJ...146..480J}
}

@ARTICLE{1978ApJ...221L..29B,
       author = {{Blandford}, R.~D. and {Ostriker}, J.~P.},
        title = "{Particle acceleration by astrophysical shocks.}",
      journal = {\apjl},
         year = 1978,
        month = apr,
       volume = {221},
        pages = {L29-L32},
          doi = {10.1086/182658},
       adsurl = {https://ui.adsabs.harvard.edu/abs/1978ApJ...221L..29B}
}

@ARTICLE{1978MNRAS.182..147B,
       author = {{Bell}, A.~R.},
        title = "{The acceleration of cosmic rays in shock fronts - I.}",
      journal = {\mnras},
         year = 1978,
        month = jan,
       volume = {182},
        pages = {147-156},
          doi = {10.1093/mnras/182.2.147},
       adsurl = {https://ui.adsabs.harvard.edu/abs/1978MNRAS.182..147B}
}

@ARTICLE{1970ApJ...162.1049H,
       author = {{Hasselmann}, K. and {Wibberenz}, G.},
        title = "{A Note on the Parallel Diffusion Coefficient}",
      journal = {\apj},
         year = 1970,
        month = dec,
       volume = {162},
        pages = {1049},
          doi = {10.1086/150736},
       adsurl = {https://ui.adsabs.harvard.edu/abs/1970ApJ...162.1049H}
}

@ARTICLE{2024ApJ...965...61C,
       author = {{Chen}, Xiaohang and {Giacalone}, Joe and {Guo}, Fan and {Klein}, Kristopher G.},
        title = "{Parallel Diffusion Coefficient of Energetic Charged Particles in the Inner Heliosphere from the Turbulent Magnetic Fields Measured by Parker Solar Probe}",
      journal = {\apj},
         year = 2024,
        month = apr,
       volume = {965},
       number = {1},
          eid = {61},
        pages = {61},
          doi = {10.3847/1538-4357/ad33c3},
archivePrefix = {arXiv},
       eprint = {2403.08141},
 primaryClass = {astro-ph.SR},
       adsurl = {https://ui.adsabs.harvard.edu/abs/2024ApJ...965...61C}
}

@ARTICLE{2000JGR...10525079Z,
       author = {{Zank}, G.~P. and {Rice}, W.~K.~M. and {Wu}, C.~C.},
        title = "{Particle acceleration and coronal mass ejection driven shocks: A theoretical model}",
      journal = {\jgr},
         year = 2000,
        month = nov,
       volume = {105},
       number = {A11},
        pages = {25079-25096},
          doi = {10.1029/1999JA000455},
       adsurl = {https://ui.adsabs.harvard.edu/abs/2000JGR...10525079Z}
}

@ARTICLE{2012AdSpR..49.1067L,
       author = {{Li}, G. and {Shalchi}, A. and {Ao}, X. and {Zank}, G. and {Verkhoglyadova}, O.~P.},
        title = "{Particle acceleration and transport at an oblique CME-driven shock}",
      journal = {Advances in Space Research},
         year = 2012,
        month = mar,
       volume = {49},
       number = {6},
        pages = {1067-1075},
          doi = {10.1016/j.asr.2011.12.027},
       adsurl = {https://ui.adsabs.harvard.edu/abs/2012AdSpR..49.1067L}
}

@ARTICLE{2017JGRA..12210938H,
       author = {{Hu}, Junxiang and {Li}, Gang and {Ao}, Xianzhi and {Zank}, Gary P. and {Verkhoglyadova}, Olga},
        title = "{Modeling Particle Acceleration and Transport at a 2-D CME-Driven Shock}",
      journal = {Journal of Geophysical Research (Space Physics)},
         year = 2017,
        month = nov,
       volume = {122},
       number = {11},
        pages = {10,938-10,963},
          doi = {10.1002/2017JA024077},
       adsurl = {https://ui.adsabs.harvard.edu/abs/2017JGRA..12210938H}
}

@ARTICLE{2025JGRA..13033147K,
       author = {{Kinoshita}, G. and {Ueno}, H. and {Murakami}, G. and {Pinto}, M. and {Yoshioka}, K. and {Miyoshi}, Y.},
        title = "{Simulation for the Calibration of Radiation Housekeeping Monitor Onboard BepiColombo/MMO and Application to the Inner Heliosphere Exploration}",
      journal = {Journal of Geophysical Research (Space Physics)},
         year = 2025,
        month = jan,
       volume = {130},
       number = {1},
        pages = {2024JA033147},
          doi = {10.1029/2024JA033147},
       adsurl = {https://ui.adsabs.harvard.edu/abs/2025JGRA..13033147K}
}

@INPROCEEDINGS{2021GMS...262...63T,
       author = {{Townsend}, Lawrence W.},
        title = "{Effects of Space Radiation on Humans in Space Flight}",
    booktitle = {Space Weather Effects and Applications},
         year = 2021,
       editor = {{Coster}, Anthea J. and {Erickson}, Philip J. and {Lanzerotti}, Louis J. and {Zhang}, Yongliang and {Paxton}, Larry J.},
       volume = {5},
        month = apr,
        pages = {63},
          doi = {10.1002/9781119815570.ch3},
       adsurl = {https://ui.adsabs.harvard.edu/abs/2021GMS...262...63T}
}

@INPROCEEDINGS{2021GMS...262...79M,
       author = {{Mertens}, Christopher J. and {Tobiska}, W. Kent},
        title = "{Space Weather Radiation Effects on High-Altitude/-Latitude Aircraft}",
    booktitle = {Space Weather Effects and Applications},
         year = 2021,
       editor = {{Coster}, Anthea J. and {Erickson}, Philip J. and {Lanzerotti}, Louis J. and {Zhang}, Yongliang and {Paxton}, Larry J.},
       volume = {5},
        month = apr,
        pages = {79},
          doi = {10.1002/9781119815570.ch4},
       adsurl = {https://ui.adsabs.harvard.edu/abs/2021GMS...262...79M}
}

@ARTICLE{2021SpWea..1902750K,
       author = {{Kress}, B.~T. and {Rodriguez}, J.~V. and {Boudouridis}, A. and {Onsager}, T.~G. and {Dichter}, B.~K. and {Galica}, G.~E. and {Tsui}, S.},
        title = "{Observations From NOAA's Newest Solar Proton Sensor}",
      journal = {Space Weather},
         year = 2021,
        month = dec,
       volume = {19},
       number = {12},
          eid = {e02750},
        pages = {e02750},
          doi = {10.1029/2021SW002750},
       adsurl = {https://ui.adsabs.harvard.edu/abs/2021SpWea..1902750K}
}

@article{kitagawa1996monte,
  title={Monte Carlo filter and smoother for non-Gaussian nonlinear state space models},
  author={Kitagawa, Genshiro},
  journal={Journal of computational and graphical statistics},
  volume={5},
  number={1},
  pages={1--25},
  year={1996},
  publisher={Taylor \& Francis}
}

@inproceedings{gordon1993novel,
  title={Novel approach to nonlinear/non-Gaussian Bayesian state estimation},
  author={Gordon, Neil J and Salmond, David J and Smith, Adrian FM},
  booktitle={IEE proceedings F (radar and signal processing)},
  volume={140},
  number={2},
  pages={107--113},
  year={1993},
  organization={IET}
}

@ARTICLE{2009EM&P..104..295G,
       author = {{Gopalswamy}, N. and {Yashiro}, S. and {Michalek}, G. and {Stenborg}, G. and {Vourlidas}, A. and {Freeland}, S. and {Howard}, R.},
        title = "{The SOHO/LASCO CME Catalog}",
      journal = {Earth Moon and Planets},
         year = 2009,
        month = apr,
       volume = {104},
       number = {1-4},
        pages = {295-313},
          doi = {10.1007/s11038-008-9282-7},
       adsurl = {https://ui.adsabs.harvard.edu/abs/2009EM&P..104..295G}
}

@ARTICLE{2008SSRv..136..117L,
       author = {{Luhmann}, J.~G. and {Curtis}, D.~W. and {Schroeder}, P. and {McCauley}, J. and {Lin}, R.~P. and {Larson}, D.~E. and {Bale}, S.~D. and {Sauvaud}, J. -A. and {Aoustin}, C. and {Mewaldt}, R.~A. and {Cummings}, A.~C. and {Stone}, E.~C. and {Davis}, A.~J. and {Cook}, W.~R. and {Kecman}, B. and {Wiedenbeck}, M.~E. and {von Rosenvinge}, T. and {Acuna}, M.~H. and {Reichenthal}, L.~S. and {Shuman}, S. and {Wortman}, K.~A. and {Reames}, D.~V. and {Mueller-Mellin}, R. and {Kunow}, H. and {Mason}, G.~M. and {Walpole}, P. and {Korth}, A. and {Sanderson}, T.~R. and {Russell}, C.~T. and {Gosling}, J.~T.},
        title = "{STEREO IMPACT Investigation Goals, Measurements, and Data Products Overview}",
      journal = {\ssr},
         year = 2008,
        month = apr,
       volume = {136},
       number = {1-4},
        pages = {117-184},
          doi = {10.1007/s11214-007-9170-x},
       adsurl = {https://ui.adsabs.harvard.edu/abs/2008SSRv..136..117L}
}

@ARTICLE{2008SSRv..136..437G,
       author = {{Galvin}, A.~B. and {Kistler}, L.~M. and {Popecki}, M.~A. and {Farrugia}, C.~J. and {Simunac}, K.~D.~C. and {Ellis}, L. and {M{\"o}bius}, E. and {Lee}, M.~A. and {Boehm}, M. and {Carroll}, J. and {Crawshaw}, A. and {Conti}, M. and {Demaine}, P. and {Ellis}, S. and {Gaidos}, J.~A. and {Googins}, J. and {Granoff}, M. and {Gustafson}, A. and {Heirtzler}, D. and {King}, B. and {Knauss}, U. and {Levasseur}, J. and {Longworth}, S. and {Singer}, K. and {Turco}, S. and {Vachon}, P. and {Vosbury}, M. and {Widholm}, M. and {Blush}, L.~M. and {Karrer}, R. and {Bochsler}, P. and {Daoudi}, H. and {Etter}, A. and {Fischer}, J. and {Jost}, J. and {Opitz}, A. and {Sigrist}, M. and {Wurz}, P. and {Klecker}, B. and {Ertl}, M. and {Seidenschwang}, E. and {Wimmer-Schweingruber}, R.~F. and {Koeten}, M. and {Thompson}, B. and {Steinfeld}, D.},
        title = "{The Plasma and Suprathermal Ion Composition (PLASTIC) Investigation on the STEREO Observatories}",
      journal = {\ssr},
         year = 2008,
        month = apr,
       volume = {136},
       number = {1-4},
        pages = {437-486},
          doi = {10.1007/s11214-007-9296-x},
       adsurl = {https://ui.adsabs.harvard.edu/abs/2008SSRv..136..437G}
}

@ARTICLE{2024Sci...385..962R,
       author = {{Rivera}, Yeimy J. and {Badman}, Samuel T. and {Stevens}, Michael L. and {Verniero}, Jaye L. and {Stawarz}, Julia E. and {Shi}, Chen and {Raines}, Jim M. and {Paulson}, Kristoff W. and {Owen}, Christopher J. and {Niembro}, Tatiana and {Louarn}, Philippe and {Livi}, Stefano A. and {Lepri}, Susan T. and {Kasper}, Justin C. and {Horbury}, Timothy S. and {Halekas}, Jasper S. and {Dewey}, Ryan M. and {De Marco}, Rossana and {Bale}, Stuart D.},
        title = "{In situ observations of large-amplitude Alfv{\'e}n waves heating and accelerating the solar wind}",
      journal = {Science},
         year = 2024,
        month = aug,
       volume = {385},
       number = {6712},
        pages = {962-966},
          doi = {10.1126/science.adk6953},
archivePrefix = {arXiv},
       eprint = {2409.00267},
 primaryClass = {astro-ph.SR},
       adsurl = {https://ui.adsabs.harvard.edu/abs/2024Sci...385..962R}
}

@ARTICLE{2020SSRv..216..125B,
       author = {{Baumjohann}, W. and {Matsuoka}, A. and {Narita}, Y. and {Magnes}, W. and {Heyner}, D. and {Glassmeier}, K.-H. and {Nakamura}, R. and {Fischer}, D. and {Plaschke}, F. and {Volwerk}, M. and {Zhang}, T.~L. and {Auster}, H.-U. and {Richter}, I. and {Balogh}, A. and {Carr}, C.~M. and {Dougherty}, M. and {Horbury}, T.~S. and {Tsunakawa}, H. and {Matsushima}, M. and {Shinohara}, M. and {Shibuya}, H. and {Nakagawa}, T. and {Hoshino}, M. and {Tanaka}, Y. and {Anderson}, B.~J. and {Russell}, C.~T. and {Motschmann}, U. and {Takahashi}, F. and {Fujimoto}, A.},
        title = "{The BepiColombo-Mio Magnetometer en Route to Mercury}",
      journal = {\ssr},
         year = 2020,
        month = oct,
       volume = {216},
       number = {8},
          eid = {125},
        pages = {125},
          doi = {10.1007/s11214-020-00754-y},
       adsurl = {https://ui.adsabs.harvard.edu/abs/2020SSRv..216..125B}
}

\clearpage

\clearpage

\clearpage

\clearpage

\clearpage

\clearpage

\clearpage

\clearpage

\clearpage

\end{document}